\documentclass[a4paper,11pt]{article}
\pdfoutput=1 
\usepackage{jhep_like}

\usepackage{amsmath,amssymb,color,epsf}
\usepackage{graphicx}
\usepackage{subfloat}
\usepackage{subfig}
\usepackage{multirow}
\usepackage{ulem}
\usepackage{comment}
\usepackage{appendix}
\usepackage{tikz}

\newcommand{\be}{\begin{equation}}
\newcommand{\ee}{\end{equation}}
\newcommand{\bea}{\setlength\arraycolsep{2pt} \begin{eqnarray}}
\newcommand{\eea}{\end{eqnarray}}
\newcommand{\nn}{\nonumber}

\def\0{{\sst{(0)}}}
\def\1{{\sst{(1)}}}
\def\2{{\sst{(2)}}}
\def\3{{\sst{(3)}}}
\def\4{{\sst{(4)}}}
\def\5{{\sst{(5)}}}
\def\6{{\sst{(6)}}}
\def\7{{\sst{(7)}}}
\def\8{{\sst{(8)}}}
\def\sst#1{{\scriptscriptstyle #1}}

\def\avg#1{\left\langle#1\right\rangle}

\def\Re{{\rm Re}}

\def\be{\begin{equation}}       \def\ee{\end{equation}}
\def\bea{\begin{eqnarray}}      \def\eea{\end{eqnarray}}
\def\ba{\begin{array}}
	\def\ea{\end{array}}
\def\bnum{\begin{enumerate} }
	\def\enum{\end{enumerate}}

\def\nn{\nonumber}

\def\=>{\Rightarrow}
\def\>{\rightarrow}

\def\eye2{Fathbb{I}}

\def\Re{\mathrm{Re}}

\def\Tr{\mathrm{Tr}}

\def\tr{\mathrm{Tr}}

\def\Arg{\mathrm{Arg}}
\def\nn{\nonumber}

\title{\boldmath Detecting quantum chaos via pseudo-entropy}

\author[a,b,c,d]{Song He,}
\author[e,f,g]{Pak Hang Chris Lau,}
\author[a]{and Long Zhao\footnote{Corresponding author}}

\affiliation[a]{Center for Theoretical Physics and College of Physics, Jilin University,  \newline Changchun 130012, People's Republic of China}
\affiliation[b]{Institute of Fundamental Physics and Quantum Technology, \newline Ningbo University, Ningbo, Zhejiang 315211, China}
\affiliation[c]{Max Planck Institute for Gravitational Physics (Albert Einstein Institute),  \newline Am M\"uhlenberg 1, 14476 Golm, Germany}
\affiliation[d]{School of Physical Science and Technology, Ningbo University, Ningbo, 315211, China}
\affiliation[e]{School of Sciences, Great Bay University, Dongguan 523000, China}
\affiliation[f]{Department of Physics, Osaka University, Toyonaka, Osaka 560-0043, Japan.}
\affiliation[g]{Department of Physics, Kobe University, Kobe-shi, Hyogo 657-8501, Japan}

\emailAdd{hesong@nbu.com}
\emailAdd{phcl2@gbu.edu.cn}
\emailAdd{zhaolong@jlu.edu.cn}

\preprint{ }

\abstract{Quantum informatic quantities such as entanglement entropy are useful in detecting quantum phase transitions. Recently, a new entanglement measure called pseudo-entropy was proposed which is a generalization of the more well-known entanglement entropy. It has many nice properties and is useful in the study of post-selection measurements. In this paper, one of our goals is to explore the properties of pseudo-entropy and study its effectiveness as a quantum chaos diagnostic, i.e., as a tool to distinguish between chaotic and integrable systems. Using various variants of the SYK model, we study the signal of quantum chaos captured in the pseudo-entropy and relate it to the spectral form factor (SFF) and local operator entanglement (LOE).}

\begin{document}

\maketitle

\section{Introduction}
\label{sec:intro}

The authors in~\cite{Hayden:2007cs,Sekino:2008he} posit that a quantum system suitable for describing a black hole must exhibit quantum chaotic behavior, based on certain general assumptions. This notion finds support in the anti-de Sitter/conformal field theory (AdS/CFT) correspondence~\cite{Maldacena:1997re,Gubser:1998bc,Witten:1998qj}. According to AdS/CFT, the boundary dual of a black hole geometry in a $d$-dimensional asymptotically AdS spacetime corresponds to a CFT in a thermofield double (TFD) state in $d-1$ dimension. By examining the out-of-time-order (OTOC) correlation function in the CFT, one can identify a non-zero Lyapunov exponent, which indicates the system is chaotic~\cite{Shenker:2013pqa, Shenker:2014cwa}.

According to the discussion in Ref.~\cite{Hosur:2015ylk}, the OTOC can be used to measure the change of quantum entanglement between a subsystem and its complement. Therefore, one can diagnose quantum chaos by entanglement entropy. Quantum entanglement is another crucial aspect of quantum gravity~\cite{VanRaamsdonk:2010pw, Maldacena:2013xja}. Ryu and Takayanagi proposed that the entanglement entropy of a subregion in the boundary conformal field theory is equivalent to the area of the corresponding minimal surface in the bulk. This connection is known as the holographic entanglement entropy conjecture~\cite{Ryu:2006bv,Ryu:2006ef} and is an essential component of the AdS/CFT correspondence framework \cite{Liu:2013qca,Liu:2013iza,Fan:2016ean,Mezei:2016wfz}. Besides OTOC, there are also other quantities that serve as probes of quantum chaos, such as the K-complexity~\cite{Parker:2018yvk,Jian:2020qpp,Rabinovici:2020ryf,Dymarsky:2021bjq,Liu:2022god,Adhikari:2022whf,Rabinovici:2022beu,Bhattacharyya:2023dhp,Bhattacharyya:2023grv,Li:2024kfm}, the energy level spacing distribution, and the Spectral Form Factor (SFF)~\cite{Garcia-Garcia:2016mno,Cotler:2016fpe,Li:2017hdt,Liu:2018hlr,deMelloKoch:2019rxr,Ma:2020uox,Bhattacharyya:2023gvg}, etc. Therefore, it is also interesting to establish connections between these quantities and quantum entanglement.

The pseudo-entropy is a generalization of the entanglement entropy in the sense that entanglement entropy appears as a special case. Entanglement entropy captures the quantum entanglement between a subregion ($A$) and its complement ($\bar{A}$) of a pure state. Given a pure state $\rho=\left| \psi \right\rangle \left\langle \psi \right|$ of a system, the entanglement entropy is defined as the von Neumann entropy of the reduced density matrix $\rho_A=\tr_{\bar{A}} \rho$. 
\begin{align}
    S_A = -\tr \rho_A \log \rho_A \,.
\end{align}
For the case of pseudo-entropy, a transition matrix between two states $\left| \psi \right\rangle$ and $\left| \varphi \right\rangle$ is constructed instead of the usual density matrix
\begin{align}
\mathcal{T}^{\psi|\varphi}=\frac{|\psi\rangle\langle\varphi|}{\langle\varphi|\psi\rangle} \,,
\end{align}
where the transition matrix is generally non-hermitian, and its trace is normalized to unity\footnote{
In addition to pure-state transition matrices, one may also attempt to define mixed-state transition matrices. The entanglement between subsystems $A$ and $\bar{A}$ in a mixed state $\rho$ can be quantified by the negativity~\cite{Bennett:1995ra,Peres:1996dw,Calabrese:2012ew,Kusuki:2019zsp}, defined as the sum of absolute values of negative eigenvalues obtained after performing a partial transpose on $\rho$ with respect to subsystem $A$. When the partially transposed $\rho$ has no negative eigenvalues, the subsystems are unentangled. However, the reduced transition matrix obtained by taking the partial trace of a pure-state transition matrix is an arbitrary matrix with complex eigenvalues. Consequently, the presence of negative eigenvalues after partial transposition cannot determine whether quantum entanglement exists between $A$ and $\bar{A}$. In fact, a pure-state transition matrix (a non-Hermitian matrix) has eigenvalues restricted to 0 and 1 and can be proven to be pseudo-Hermitian. An $\eta$-pseudo-Hermitian matrix $M$ satisfies:
\begin{align}
\eta M\eta^{-1}=M^\dagger\,,
\end{align}
where $\eta$ is an invertible Hermitian operator. Such matrices have either real eigenvalues or complex conjugate pairs. Crucially, the product of two Hermitian matrices is $\eta$-pseudo-Hermitian~\cite{Mostafazadeh:2001jk,Guo:2022jzs}. If both Hermitian matrices are positive semi-definite, their product has exclusively non-negative real eigenvalues~\cite{He:2023syy}. This motivates the following definition of a mixed-state transition matrix: by choosing the initial state as density matrix $\rho_2$ and the final state as $\rho_1$, the transition matrix becomes~\cite{Guo:2022jzs}
\begin{align}
\mathcal{T}=\frac{\rho_1\rho_2}{\tr[\rho_1\rho_2]}\,.
\label{eq:TM-mixed}
\end{align}
When $\rho_1$ and $\rho_2$ are pure-state density matrices, $\mathcal{T}$ reduces to the pure-state transition matrix. Notably, $\mathcal{T}$ is $\eta$-pseudo-Hermitian with non-negative real eigenvalues. This constraint provides a viable pathway to generalize negativity to transition matrices.}.

The computation of the pseudo-entropy closely resembles its entanglement entropy counterpart. We first partial trace over a subregion of the transition matrix to obtain the reduced transition matrix
\begin{align}
\mathcal{T}_A^{\psi|\varphi}=\tr_{\bar{A}}\mathcal{T}^{\psi|\varphi}\,.
\end{align}
Then the pseudo-entropy is obtained by substituting this reduced transition matrix into the von Neumann entropy formula
\begin{align}
    S(\mathcal{T}_A^{\psi|\varphi})=-\tr_A\mathcal{T}_A^{\psi|\varphi}\log\mathcal{T}_A^{\psi|\varphi}\ \,.
\end{align}
Alternatively, it can be computed by the replica trick with the help of a generalized version of the R\'enyi entropy. The $n$-th pseudo-R\'enyi entropy of the transition matrix $\mathcal{T}_A^{\psi|\varphi}$ is defined as 
\begin{align}
    S^{(n)}(\mathcal{T}_A^{\psi|\varphi})
    =\frac{1}{1-n}\log\tr\left[\left(\mathcal{T}_A^{\psi|\varphi}\right)^n\right] \,,
\end{align}
where the pseudo-entropy is recovered by taking the $n\rightarrow 1$ limit. There is also substantial research based on the pseudo-entropy~\cite{Nakata:2020luh,Mollabashi:2020yie,Mollabashi:2021xsd} and its connection to time-like entanglement~\cite{Doi:2022iyj, Li:2022tsv, Doi:2023zaf,Jiang:2023loq,Jiang:2023ffu,Chen:2023eic}, the information paradox through the study of Page curve~\cite{Akal:2021dqt} and 2-dimensional CFT where analytic computation is possible~\cite{Guo:2022sfl,Guo:2022jzs,He:2023eap,He:2023wko}. The mathematical structure of transition matrices is more general than that of density matrices. Therefore, pseudo-entropy can be more generally connected to other quantities when compared with the entanglement entropy. The main goal of this paper is to explore further the properties of pseudo-entropy and its application to quantum chaos based on the work~\cite{Goto:2021kln}, where the authors demonstrated that the connection of the pseudo-entropy to the spectral form factor. Although the real part of pseudo-entropy has been extensively studied, the physical significance of its imaginary part remains an open question under investigation. In a recent work~\cite{Caputa:2024gve}, Caputa et al. has discussed that, for transition matrices like Eq.~\eqref{eq:TM-TFD}, the imaginary part of the pseudo-entropy can be derived from its real part via the Kramers-Kronig relation.

In this paper, we will focus on the sparse SYK model, which was first proposed in~\cite{Xu:2020shn, Garcia-Garcia:2020cdo}. It is a modification of the original SYK model which contains an interesting feature of chaos/integrable transition. There have been many works studying this modified model through the use of spectral form factor~\cite{Caceres:2022kyr} and OTOC~\cite{Caceres:2023yoj}. Besides as a model of quantum chaos, it has also been used as a simple model for modeling a transversable wormhole~\cite{Caceres:2021nsa, Jafferis:2022crx}. One can further simplify the sparse SYK model by restricting the allowed value of the coupling constants to $\pm1$ and it becomes the binary sparse model~\cite{Tezuka:2022mrr}. The advantage of this much-simplified model is that it retains the essential features of the original model but is more numerically efficient. We will also take advantage of it and study the properties of pseudo-entropy using this model.

The structure of this paper is as follows. In Section~\ref{sec:SYK}, we give an overview of the SYK model and its variants. In Section~\ref{sec:SFF}, we derive the relationship between the SFF and the pseudo-R\'enyi entropy. In Section~\ref{sec:LOE}, we establish the relationship between the pseudo-R\'enyi entropy and the OTOC. We end in Section~\ref{Sec:Summary} with a summary and an outlook.

\section{Review of the SYK model and its variants}
\label{sec:SYK}
The SYK model is a $q$-body all-to-all interacting theory of $N$ Majorana fermions. Its full Hamiltonian is~\cite{Kit.KITP.1,Kit.KITP,Maldacena:2016hyu}
\begin{align}
    H=i^{q/2}\sum_{1\leq i_1<\cdots<i_q\leq N}J_{i_1\cdots i_q}\psi_{i_1}\cdots\psi_{i_q}\,,
    \label{eq:Hamiltonian-SYK}
\end{align}
where the Majorana fermions, $\psi$, satisfy the anticommutation relation $\{\psi_i,\psi_j\}=2\delta_{ij}$, the coupling tensor $J_{j_1j_2\cdots j_q}$ is totally antisymmetric, and each independent element is randomly drawn from a Gaussian distribution with zero mean and variance $\avg{J_{j_1j_2\cdots j_q}^2}=\frac{(q-1)!}{N^{q-1}}J_0^2=\frac{2^{q-1} (q-1)! }{q N^{q-1}}\mathcal{J}^2$. 

The spectral statistics of the SYK model can be effectively described by Random Matrix Theory (RMT). Specifically, in the case where $q$ modulo 4 is equal to 0, the spectral statistics fall under the Gaussian Orthogonal Ensemble (GOE) for $N$ modulo 8 equal to 0, the Gaussian Symplectic Ensemble (GSE) for $N$ modulo 8 equal to 4, and the Gaussian Unitary Ensemble (GUE) for $N$ modulo 8 equal to 2 or 6~\cite{Cotler:2016fpe}. On the other hand, in the case where $q$ modulo 4 is equal to 2, the spectral statistics belong to Class C (BdG) for $N$ modulo 8 equal to 4 and Class D (BdG) for $N$ modulo 8 equal to 0~\cite{Li:2017hdt,Kanazawa:2017dpd,Sun:2019yqp}.

Recently, a sparse version of the SYK model was proposed \cite{Xu:2020shn}. The full Hamiltonian of the sparse SYK$_q$ model with $N$ Majorana fermions \cite{Xu:2020shn} is given by
\begin{align}
H = i^{q/2}\sum_{1\leq i_1<\cdots<i_q\leq N}x_{i_1\cdots i_q}J_{i_1\cdots i_q}\psi_{i_1}\cdots\psi_{i_q} \,,
\label{eq:Hamiltonian-sparseSYK}
\end{align}
where the parameters $x_{i_1\cdots i_q}$ are chosen to be 1 with probability $p$ and 0 with probability $1-p$. The number of non-zero terms in the sum (\ref{eq:Hamiltonian-sparseSYK}) is approximately given by
\begin{align}
k_{cpl} = p\binom{N}{q} \approx \frac{pN^q}{q!} \, .
\end{align}
The amount of the sparseness of the model is controlled by the parameter 
\begin{align}
    k=k_{cpl}/N\approx\frac{pN^{q-1}}{q!}\,.
\end{align}
This sparse version of the SYK model has been proven that it exhibits chaotic behavior when $k$ exceeds a certain $\mathcal{O}(1)$ constant \cite{Xu:2020shn, Garcia-Garcia:2020cdo}. The coupling constants $J_{i_1\cdots i_q}$ in Eq.~(\ref{eq:Hamiltonian-sparseSYK}) are Gaussian random variables with zero mean and variance
\begin{align}
    \langle J_{i_1i_2\cdots i_q}^2\rangle_J=\frac{(q-1)!J^2}{N^{q-1}}\frac{1}{p}\,,
\end{align}
where the factor $p$ in the denominator compensates for the dependence of the average energy on $p$. The spectral statistics of the sparse SYK model with $q=4$ have been discussed in \cite{Caceres:2022kyr, Tezuka:2022mrr}. For sufficiently large $k$, the sparse SYK model can be classified into GOE, GSE, and GUE as in the original SYK model. However, for small values of $k$ or $k_{cpl}$, the energy gap ratio is about 0.386 \cite{Tezuka:2022mrr}, indicating that the energy level spacing distribution of the sparse SYK model follows the Poisson distribution \cite{PhysRevLett.110.084101}.

The Majorana fermions in the SYK model can be expressed as a spin chain using the Jordan-Wigner transformation. The convention used in this paper is
\begin{align}
    \psi_{2i-1}&= \underbrace{\sigma_z \otimes \cdots \otimes \sigma_z}_{i-1} \otimes \ \sigma_x \otimes \underbrace{ I \otimes \cdots \otimes I }_{\frac{N}{2}-i} \,, \label{eq:psiodd-JW-SYK}\\
    \psi_{2i} &=  \underbrace{\sigma_z \otimes \cdots \otimes \sigma_z}_{i-1} \otimes \ \sigma_y \otimes \underbrace{ I \otimes \cdots \otimes I }_{\frac{N}{2}-i} \,,
    \label{eq:psieven-JW-SYK}
\end{align}
where $i=1,\cdots,\frac{N}{2}$ with $N$ the number of Majorana fermions in consideration. On this basis, the Majorana fermions are viewed as a non-local object. In this paper, the partial trace operations are applied to a subset of qubits on this basis when computing the pseudo-entropy numerically. One potential problem of this choice is the difficulty in physically interpreting such operations in the original Majorana fermion basis. 

An alternative SYK-like model was recently proposed by formulating a randomly coupled model using qubits instead of fermions, and it is called the SpinXY$_4$ model \cite{Hanada:2023rkf}. It closely resembles and demonstrates the chaotic behaviors of the SYK model. The SpinXY$_4$ model Hamiltonian with $\frac{N_{\text{Maj}}}{2}$ qubits, where $N_{\text{Maj}}$ is the number of Majorana fermion which gives the same Hilbert space dimension, is given by
\begin{align}
    H_{\text{Spin}}&= \sqrt{\frac{6}{N_{\text{Maj}}^3}} \sum_{1 \leq a<b<c<d\leq N_{\text{Maj}}} J_{abcd} i^{\eta_{abcd}} {\cal O}_{a} {\cal O}_{b}{\cal O}_{c}{\cal O}_{d} \,,
\end{align}
where the spin operators ${\cal O}$ are
\begin{align}
    {\cal O}_{2i-1} &= \underbrace{I \otimes \cdots \otimes I}_{i-1} \otimes \ \sigma_x \otimes \underbrace{ I \otimes \cdots \otimes I }_{\frac{N}{2}-i} \,,\\
    {\cal O}_{2i} &= \underbrace{I \otimes \cdots \otimes I}_{i-1} \otimes \ \sigma_y \otimes \underbrace{ I \otimes \cdots \otimes I }_{\frac{N}{2}-i} \,.
    \label{eq:def-PauliXY}
\end{align}
The coupling constants follow a probability distribution
\begin{align}
    P(J_{abcd}) &= \frac{1}{\sqrt{2\pi}} e^{-J_{abcd}^2/2} \,,
\end{align}
which corresponds to zero mean and a variance of unity with this choice of scaling. The extra factor of $i^{\eta_{abcd}}$ comes from the fact that the ${\cal O}_i$ operators are effectively a Pauli operator at a single site (qubit). Due to the algebra of the Pauli operators, if both ${\cal O}_{2i-1} {\cal O}_{2i}$ appear in a single term of the Hamiltonian. The Hamiltonian will no longer be Hermitian unless a factor of $i$ is inserted by hand. The factor $\eta_{abcd}$ is defined such that the Hermiticity of the Hamiltonian recovered. It counts the number of ${\cal O}_{2i-1} {\cal O}_{2i}$ pairs in the term e.g. $\eta_{1237}=1$ with ${\cal O}_{1} {\cal O}_{2}$ as a pair and $\eta_{1256}=2$ with ${\cal O}_{1} {\cal O}_{2}$ and ${\cal O}_{5} {\cal O}_{6}$ as two pairs. One advantage of this model is that these operators are local in the spin basis and as a result, the partial trace/transpose operations are better physically interpreted. We will also use this model as an example of a chaotic system to study the properties of pseudo-entropy in this paper.

\section{Pseudo-entropy as a probe of chaos}
\label{sec:SFF}
Before delving into Section 3.1, let's recap the main focus of this chapter. The core of this chapter is to explore the manifestation of quantum chaos through pseudo-entropy. As an emerging quantum information metric, pseudo-entropy offers broader applications compared to traditional entanglement entropy. By studying various variants of the SYK model, we investigate the effectiveness of pseudo-entropy in distinguishing chaotic systems from integrable systems.

\subsection{Chaotic spectrum via pseudo-entropy}
\label{sec:SFF-PEE}

The spectral information of a disordered system is captured in the spectral form factor (SFF, denoted as $g(t,\beta)$). It is given by ~\cite{Garcia-Garcia:2016mno,Cotler:2016fpe,Cotler:2017jue}
\begin{align} 
	g(t, \beta) &=\frac{\left\langle Z(\beta +it) Z^{*}(\beta+it)\right\rangle_{J}}{\left\langle Z(\beta)\right\rangle_{J}^{2}} \,,
    \label{eqn:sff}
\end{align}
where $Z(\beta+it)=\tr \, e^{-(\beta+it)H}$ is the partition function with a complex temperature. It is sometimes more useful to subtract the disconnected pieces, $g_d(t,\beta)$, of the SFF to review the true correlation of the spectrum. The resulting quantity is the connected SFF, $g_c(t,\beta)$, given by
\begin{align}
	g_{d}(t, \beta) &=\frac{\langle Z(\beta+i t)\rangle_{J} \cdot\left\langle Z^{*}(\beta +it)\right\rangle_{J}}{\langle Z(\beta)\rangle_{J}^{2}} \,, \\ 
	g_{c}(t, \beta) &=g(t, \beta)-g_{d}(t, \beta) \,.
	\label{eq:SFF-definition}
\end{align}
At infinite temperature, the late time behavior of the SFF, $g(t,0)$, is solely determined by the symmetry of the theory~\cite{Cotler:2016fpe}.

To review spectral information captured in pseudo-entropy, we consider a two-sided model with two uncoupled SYK models. We consider the transition matrix connecting the following two states
\begin{align}
    &|\psi\rangle=\frac{1}{Z(\beta)^{1/2}}\sum_n e^{-\frac{\beta}{4}(H_L+H_R)}|n_L\rangle\otimes|n_R\rangle \,,\nn \\
    &|\varphi\rangle=e^{\frac{it}{2}(H_L+H_R)}|\psi\rangle\,,
\end{align}
where $\left| \psi \right\rangle$ is the thermofield double state (TFD). The TFD state can be understood as the canonical purification of the thermal density matrix of one of the two SYK models. The state $\left| \varphi \right\rangle$ can be interpreted as a one-sided time evolution of the TFD state. To compute the pseudo-entropy, we first construct the transition matrix\footnote{In defining the transition matrix, we have introduced a normalization factor given by the inner product between the two states, $\left|\psi\right\rangle$ and $\left|\varphi\right\rangle$. We have implicitly made an assumption that these two states are not orthogonal to each other, i.e. $\langle\varphi|\psi\rangle \neq0$. In this paper, our interest is in the transition matrix constructed by the thermofield double state at temperature $\beta$ $\left| \textrm{TFD}_\beta \right\rangle \equiv \sum_n e^{-\frac{\beta}{2} E_n}\left|n \right\rangle\left|n \right\rangle$ and its time evolved state $e^{i\hat{H}t} \left| \textrm{TFD}_\beta \right\rangle = \sum_n e^{-(\frac{\beta}{2}-it) E_n}\left|n \right\rangle\left|n \right\rangle$. Since the normalization is time dependent, there is a possibility that the normalization vanishes at some time $t=t_*$, rendering the transition matrix ill-defined. This indeed can happen for a generic state. While this can occur for a generic state, there is an upper bound on how quickly a state can evolve to become orthogonal to itself, known as the quantum speed limit \cite{Deffner:2017cxz}. For the TFD transition matrix, the normalization can be written as $\left\langle \textrm{TFD}_\beta \right| e^{i\hat{H}t} \left| \textrm{TFD}_\beta \right\rangle$, and its modulus squared is referred to as the surviving probability, which has been shown to be equivalent to the SFF, $\left|\left\langle \textrm{TFD}_\beta \right| e^{i\hat{H}t} \left| \textrm{TFD}_\beta \right\rangle\right|^2 = \left|\frac{Z(\beta+it)}{Z(\beta)}\right|^2 = \textrm{SFF}(\beta,t)$ \cite{delCampo:2017bzr,delCampo:2017ftn}. Eventually, it saturates at a finite non-zero value in the long-time limit, with the average saturation value given by $\frac{Z(2\beta)}{Z(\beta)^2}$. While the normalization factor could, in principle, vanish for a pair of arbitrary states, this does not occur for the TFD state, which is the focus of this paper.}
\begin{align}
    {\cal T}^{\psi|\varphi}= \frac{\left|\psi\right\rangle \left\langle \varphi \right|}{\left\langle \varphi | \psi \right\rangle} \,,
\end{align}
and then the reduced transition matrix of the $R$ system is obtained by partial tracing over the $L$ system
\begin{align}
    \mathcal{T}_R^{\psi|\varphi}
    =\tr_L\mathcal{T}^{\psi|\varphi}
    =\frac{e^{-(\beta+it)H_R}}{Z(\beta+it)} \,.
    \label{eq:TM-TFD}
\end{align}
Note that the reduced transition matrix resembles closely the thermal density matrix of a single SYK model with a complex temperature $\beta+it$. Substituting this reduced transition matrix to the pseudo-entropy formula and taking the real part, we obtain
\begin{align}
    \mathrm{Re}[S(\mathcal{T}_R^{\psi|\varphi})]
    &=-\mathrm{Re}\left[\left\langle\tr_R\mathcal{T}_R^{\psi|\varphi}\log\mathcal{T}_R^{\psi|\varphi}\right\rangle_J\right] \nn \\
    &=\left\langle\mathrm{Re}\left[(\beta+it)\langle H_R\rangle_{\beta+it}\right]\right\rangle_J
    +\frac{1}{2}\left\langle\log\left(\left|Z(\beta+it)\right|^2\right)\right\rangle_J\,.
    \label{eq:PE-systemR}
\end{align}
Here $\left\langle \cdots \right\rangle_J$ denotes the disorder averaging. Additionally, as a side note, instead of pseudo-entropy, we can also consider a generalization of the relative entropy for a transition matrix. In Appendix \ref{sec:pre-sff}, we provide an alternative relation between pseudo-relative entropy and the SFF. The relative entropy is more rigorously defined in the context of quantum field theory and worth further investigation.

\begin{figure}[!htb]
		\centering
		\includegraphics[width=0.6\linewidth]{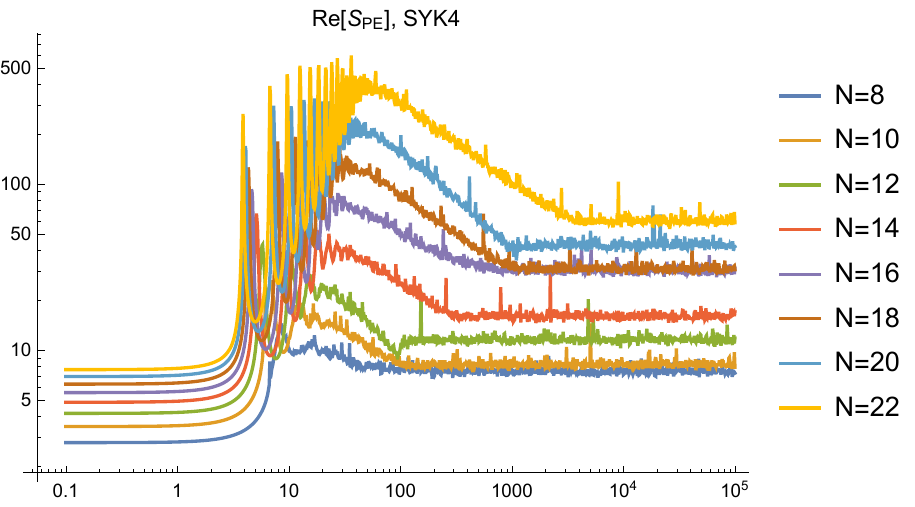}
	\caption{ The plot of the real part of the pseudo-entropy of the SYK$_4$ model with various values of $N$ and $\beta=0$.}
	\label{fig:PEE-SYK4-Ns}
\end{figure}

The result \eqref{eq:PE-systemR} has been previously discussed in the CFT$_2$ case~\cite{Goto:2021kln}. For the case of CFT$_2$, the first term can be evaluated explicitly and is proportional to $\frac{\beta}{\beta^2+t^2}$. Therefore, this term tends to zero at late times and vanishes exactly when $\beta=0$. The second term can be regarded as the logarithm of the SFF. As a result, the PEE exihibits the behavior of the SFF in their paper~\cite{Goto:2021kln}. For the SYK$_4$ model, we first calculate the pseudo-entropy numerically and display the result in Figure~\ref{fig:PEE-SYK4-Ns}. One observes that the pattern of pseudo-entropy deviates from the SFF of the SYK$_4$ model. To explain this deviation, we need to address two questions: the first question is whether the contribution of the first term in Eq.~\eqref{eq:PE-systemR} tends to zero in the $\beta = 0$ or late-time limit for the pseudo-entropy; the second question is whether the remaining part of Eq.~\eqref{eq:PE-systemR} contains information that characterizes quantum chaos. We will first discuss the second question, and then address the contribution of the first term. 

Without the disorder average, the last term in Eq.~(\ref{eq:PE-systemR}) is the logarithm of the spectral form factor (\ref{eqn:sff}) without the normalization. In the infinite temperature limit, the normalization coefficient $Z(0)^2$ in Eq.~(\ref{eq:SFF-definition}) is universal. It depends only on the Hilbert space dimension and is given by $2^{N}$ in the case of the SYK model. So without the disorder average, this term is equal to the logarithm of the spectral form factor up to an additive constant.
Assuming that we can ignore the contribution of the first term temporarily, to relate the pseudo-entropy in Eq.~\eqref{eq:PE-systemR} to the SFF, we need to express $\langle\log |Z|^2\rangle_J$ as a function of $\langle|Z|^2\rangle_J$, but computing the analytical expression of this function is challenging. We provide a numerical comparison for it in Figure.~\ref{fig:Z2-SYK4-Ns} in appendix \ref{sec:details}. We observe that as $N$ increases, $\langle\log|Z|^2\rangle_J$ and $\log\langle|Z|^2\rangle_J$ gradually converge to the same value\footnote{In Ref.~\cite{Engelhardt:2020qpv}, the authors discuss the contribution of the replica wormhole to the quenched free energy $F_{\rm ann}\equiv -T\ln\langle Z(\beta)\rangle$ based on JT gravity. They are only able to compute the contribution to the first few moments in the Airy limit. Therefore, providing a complete analytical analysis in our case is quite difficult. In our problem, we assume that the connected contributions are non-negligible, and we only discuss the approximation between $\langle\ln|Z(it)|^n\rangle$ and $\ln\langle |Z(it)|^n\rangle$.}. Thus, we assume that for large but finite $N$, we can approximately represent the SFF by the second term of Eq.~\eqref{eq:PE-systemR}. Therefore, the significant discrepancy between the results in Figure~\ref{fig:PEE-SYK4-Ns} and the SFF should primarily originate from the first term in Eq.~\eqref{eq:PE-systemR}. This contradicts the assertion in Ref.~\cite{Goto:2021kln} that the contribution of the first term is negligible. We conducted a detailed analysis of the first term's behavior in the SYK model and present the complete derivation in Appendix~\ref{sec:log}. Our investigation reveals that the dominance of the first term in Figure~\ref{fig:PEE-SYK4-Ns} stems from its involvement with complex logarithms. The logarithm of a complex number should be evaluated using the principal value of its phase angle, with its specific value depending on the location of the branch cut. Consequently, $\log e^{iHt}\neq iHt$, which implies that expanding the pseudo-entanglement entropy via the second line of Eq.~\eqref{eq:PE-systemR} is inadequate.

Based on the above discussion, we consider that the pseudo-entropy may not be capable of fully characterizing the SFF. The reason is that the definition of pseudo-entropy appears to be mathematically ambiguous. This issue is beyond the scope of this paper, and we will explore this in future work. Therefore, we shift our focus to the discussion of the relationship between pseudo-R\'enyi entropy and SFF. Since the discussion of pseudo-R\'enyi entropy only involves computing the real part of $\ln Z(it)$, the aforementioned issues do not arise.

\begin{figure}[htbp]
	\centering
	\captionsetup[subfloat]{farskip=10pt,captionskip=1pt}
	\subfloat[t][\centering{$\Re[S^{(n)}(\mathcal{T_R^{\psi|\varphi}})]$ for different $n$s}]{\label{fig:Renyi-SYK4-ns}
		\includegraphics[height =0.265\linewidth]
		{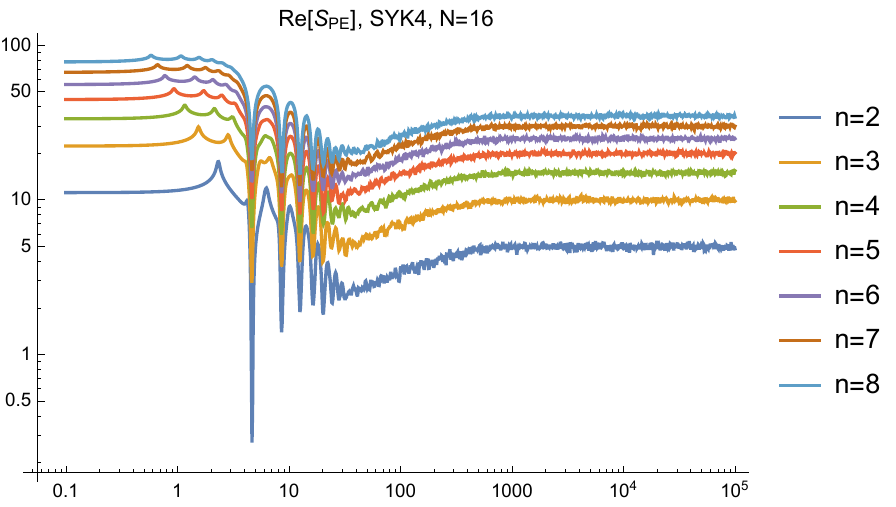}}
	\subfloat[t][\centering{$\log[R^{(2n)}(t)]$ for different $n$s}]{\label{fig:Zn-SYK4-ns}
		\includegraphics[height =0.265\linewidth]
		{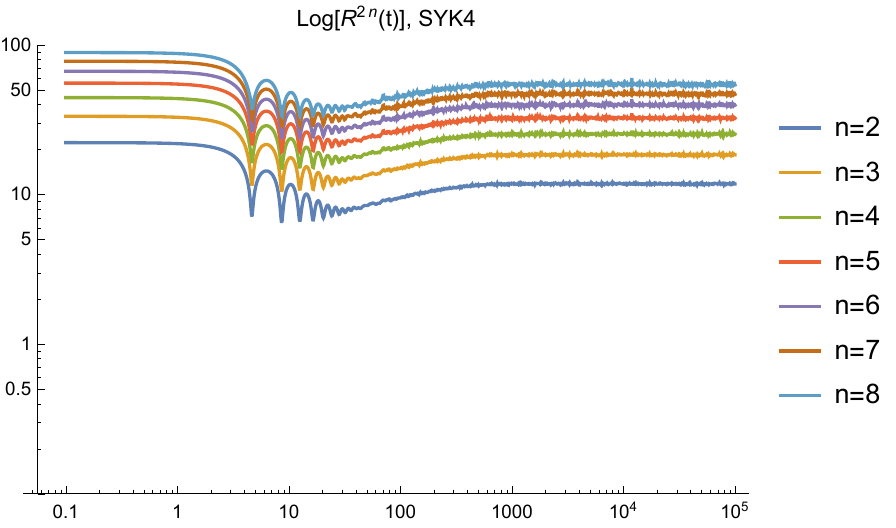}}
	\caption{(a). The real part of the $n$-th pseudo-R\'enyi entropy. (b). Logarithm of $R_{2n}(t)$. $N=16$, $n=2,\cdots 8$ and $\beta=0$.}
 \label{fig:PRE-SYK4}
	\vspace{-0.5em}
\end{figure}

The real part of the $n$-th pseudo-R\'enyi entropy of the transition matrix $\mathcal{T}_R^{\psi|\varphi}$ is given by
\begin{align}
    \Re\left[S^{(n)}(\mathcal{T}_R^{\psi|\varphi})\right]
    &=\left\langle\frac{1}{1-n}\Re\left[\log\Tr\left(\mathcal{T}_R^{\psi|\varphi}\right)^n\right]\right\rangle_J\nn\\
    &=\frac{1}{2(1-n)}\left(\left\langle\log\left|Z(n(\beta+it))\right|^2\right\rangle_J
    -\left\langle\log\left|Z(\beta+it)\right|^{2n}\right\rangle_J
    \right) \,.
    \label{eq:renyiPEE}
\end{align}
Note that the two terms in the parenthesis are not normalized, making the second term dominant in the real part of the pseudo-R\'enyi entropy for $n\gg2$. First, we still need to clarify that exchanging the order of the averaging operation and the logarithmic operation yields approximately the same result. In appendix~\ref{sec:details}, we provide the numerical comparison of the real part of the $n$-th pseudo-R\'enyi entropy, $\left\langle\log|Z|^{2n}\right\rangle_J$ and $\log\left\langle|Z|^{2n}\right\rangle_J$, for different $N$ and different $n$ respectively in Figure~\ref{fig:Renyi-Zn-SYK4-Ns} and \ref{fig:Renyi-Zn-SYK4-ks}. We observe that the curves of the $n$-th pseudo-R\'enyi entropy approach the curves of $\left\langle\log|Z(\beta+it)|^{2n}\right\rangle_J$ as $N$ and $n$ increases. We present our numerical results of $\Re\left[S^{(n)}(\mathcal{T}_R^{\psi|\varphi})\right]$ in Figure~\ref{fig:Renyi-SYK4-ns}. For comparison, we also display the logarithm of the $n$-point function in Figure~\ref{fig:Zn-SYK4-ns}. Interestingly, we observe that the time scales $t_d$ and $t_p$ are the same for all $n$, which aligns with the result in Ref.~\cite{Cotler:2016fpe}.

The behavior of the SFF at the plateau can reflect the system’s energy level degeneracy, which, in turn, reveals the system’s symmetry. We expect the pseudo-Rényi entropy to exhibit this feature as well, with analysis provided in Appendix~\ref{sec:details} and results shown in Figure~\ref{fig:normalizedRenyiPE-SYK4-Ns}. We observe that the values of $N$ mod 8 = 0 and $N$ mod 8 = 2 coincide in the plateau region, reflecting the double degeneracy of the SYK model with $N$ mod 8 = 2, 4, 6, while being nondegenerate for $N$ mod 8 = 0.

\subsection{Pseudo-entropy in the subsystem}
\label{sec:PRE-subsystem}
In Section \ref{sec:SFF-PEE}, we established that the real part of the pseudo-entropy for $\mathcal{T}_R^{\psi|\varphi}$ characterizes the properties of the spectral form factor. In this section, we focus on the pseudo-entropy of a subregion of the $R$ system. Specifically, for the SYK model, we select the first several Majorana fermions of the $R$ system as $\bar{A}$ and then trace over $\bar{A}$ to obtain the reduced transition matrix of $A$
\begin{align}
\mathcal{T}_A^{\psi|\varphi}=\tr_{\bar{A}} \mathcal{T}_R^{\psi|\varphi}\,.
\end{align}
By employing the Jordan-Wigner transformation, the state of the SYK model can be expressed in the particle number representation as follows
\begin{align}
|\Psi\rangle=\sum_{n_1,\cdots,n_{N/2}=0}^1c_{n_1,\cdots,n_{N/2}} |n_1,n_2,\cdots,n_{N/2}\rangle\,,
\end{align}
where $n_i$ represents the particle number of the $i$-th Dirac fermion in the state $|\Psi\rangle$. The eigenstate of the Hamiltonian on this basis can be decomposed as
\begin{align}
|E_n\rangle
&=\sum_{n_1,\cdots,n_{N/2}=0}^1
\langle n_1,\cdots,n_{N/2}|E_n\rangle|n_1,\cdots,n_{N/2}\rangle\nonumber\\
&=\sum_i\psi_i^n|i\rangle \,,
\label{eq:energy-wavefunction}
\end{align}
where we collectively represent the index $\{n_1,\cdots,n_{N/2}\}$ by $i$ for visual clarify and $\psi_i^n$ is the wavefunction of the $n$-th energy eigenstate in this basis. We consider a subset $A$ of the right system and trace out its complementary set $\bar{A}\cup L$. We label the quantities in subsystem $A$ and $\bar{A}$ by $N$ and $M$ respectively. The reduced transition matrix $\mathcal{T}^{\psi|\varphi}_A$ of $\mathcal{T}^{\psi|\varphi}$ is
\begin{align}
\mathcal{T}_A^{\psi|\varphi}
&=\tr_{\bar{A},L}\mathcal{T}^{\psi|\varphi}\nonumber\\
&=\frac{1}{Z(\beta+it)}\sum_{\vec{N}_1,\vec{N}_2}\sum_{\vec{M}}
\left(\langle \vec{M},\vec{N}_1|e^{-(\beta+it)H}|\vec{M},\vec{N}_2\rangle\right)
|\vec{N}_1\rangle\langle\vec{N}_2|\,,
\label{eq:induced-transition-matrix}
\end{align}
where the vector $\vec{N}/\vec{M}$ labels the eigenstates of the subsystem $A$ or $\bar{A}$ in the particle number basis. By using Eq.~(\ref{eq:energy-wavefunction}), we can express the Eq.~(\ref{eq:induced-transition-matrix}) as
\begin{align}
\mathcal{T}_A^{\psi|\varphi}
=\frac{1}{Z(\beta+it)}\sum_{\vec{N}_1,\vec{N}_2}\sum_{\vec{M}}\sum_n
(\psi_{\vec{M}\vec{N}_1}^n)^*(\psi_{\vec{M}\vec{N}_2}^n)e^{-(\beta+it)E_n}
|\vec{N}_1\rangle\langle\vec{N}_2|\,.
\label{eq:induced-transitionmatrix-numberbasis}
\end{align}

The SYK model's spectral statistics can be approximated using the Random Matrix Theory (RMT) approach, as studied in works such as \cite{Cotler:2016fpe} and \cite{Garcia-Garcia:2016mno}. For example, in the case of the Gaussian Unitary Ensemble (GUE), the spectrum statistics are invariant under unitary conjugation, leading to the following expression for the density of states
\begin{align}
dH=C|\Delta(\lambda)|^2\prod_id\lambda_idU \,,
\end{align}
where $\Delta(\lambda)$ is the Vandermonde determinant and $dU$ is the Haar measure of the unitary group $U$. The information of the eigenvector $\psi_{\vec{M}\vec{N}}^n$ are encoded in the unitary group $U$. So we can regard the eigenvector $\psi_{\vec{M}\vec{N}}^n$ of the Hamiltonian $H$ as random variables independent of the eigenvalues $E_n$. The Haar integral of the matrix elements of the unitary group element $U$ satisfies~\cite{Alhassid:2000hf, DAlessio:2015qtq, Cotler:2017jue}
\begin{align}
\int dU{U^j}_k{{U^\dagger}^l}_m=\frac{1}{L}\delta^j_m\delta^l_k\,,
\label{eq:average-unitary}
\end{align}
where $L$ is the dimension of the random matrix. Based on this relation, the disorder average of the R\'enyi entropy of the reduced transition matrix (\ref{eq:induced-transitionmatrix-numberbasis}) is computable. We will estimate the real part of the pseudo-R\'enyi entropy of the SYK model's subsystem using the RMT formula in the large-$L$ limit. Subsequently, we will complement our analysis by numerically calculating the pseudo-R\'enyi entropy for the SYK model using the exact diagonalization method.

The real part of the second pseudo-R\'enyi entropy is
\begin{align}
&-\left\langle\Re\log\tr\left(\mathcal{T}_A^{\psi|\varphi}\right)^2\right\rangle_J
=-\frac{1}{2}\left\langle\log\left|\tr\left(\mathcal{T}_A^{\psi|\varphi}\right)^2\right|^2\right\rangle_J\nonumber\\
&=-\frac{1}{2}\left\langle\log\left[
\sum_{\vec{N}_1,\vec{N}_2}\sum_{\vec{M}_1,\vec{M}_2}\sum_{n_1,n_2}
\left(\psi_{\vec{M}_1\vec{N}_1}^{n_1}\right)^*\left(\psi_{\vec{M}_1\vec{N}_2}^{n_1}\right)
\left(\psi_{\vec{M}_2\vec{N}_2}^{n_2}\right)^*\left(\psi_{\vec{M}_2\vec{N}_1}^{n_2}\right)
e^{-(\beta+it)(E_{n_1}+E_{n_2})}\right.\right.\nonumber\\
&\times\left.\left.\sum_{\vec{N}_3,\vec{N}_4}\sum_{\vec{M}_3,\vec{M}_4}\sum_{n_3,n_4}
\left(\psi_{\vec{M}_3\vec{N}_3}^{n_3}\right)\left(\psi_{\vec{M}_3\vec{N}_4}^{n_3}\right)^*
\left(\psi_{\vec{M}_4\vec{N}_4}^{n_4}\right)\left(\psi_{\vec{M}_4\vec{N}_3}^{n_4}\right)^*
e^{-(\beta-it)(E_{n_3}+E_{n_4})}\right]\right\rangle_J\nonumber\\
&+\frac{1}{2}\left\langle\log|Z(it)|^4\right\rangle_J\,.
\label{eq:2nd-renyi}
\end{align}
Up to this step, we also need to address the ensemble average of the logarithmic function of $|Z|$. { Based on the implications of the numerical results in Section~\ref{sec:SFF-PEE}, we assume that the logarithm of $\langle|Z|^2\rangle$ is well approximated by the average of $\log|Z|^2$ in the large system size limit.} Therefore, the second term becomes the logarithm of a $2n$-point function. Now let's temporarily disregard the logarithmic function and take $n=2$ to calculate the expression
\begin{align}
\sum_{\{\vec{N}_i\}}\sum_{\{\vec{M}_i\}}\sum_{\{n_i\}}
&\left\langle\left(\psi_{\vec{M}_1\vec{N}_1}^{n_1}\right)^*\left(\psi_{\vec{M}_1\vec{N}_2}^{n_1}\right)
\left(\psi_{\vec{M}_2\vec{N}_2}^{n_2}\right)^*\left(\psi_{\vec{M}_2\vec{N}_1}^{n_2}\right)\right.
\nonumber\\
&\left.\left(\psi_{\vec{M}_3\vec{N}_3}^{n_3}\right)\left(\psi_{\vec{M}_3\vec{N}_4}^{n_3}\right)^*
\left(\psi_{\vec{M}_4\vec{N}_4}^{n_4}\right)\left(\psi_{\vec{M}_4\vec{N}_3}^{n_4}\right)^*
e^{-i(E_{n_1}+E_{n_2}-E_{n_3}-E_{n_4})t}\right\rangle_J\,.
\label{eq:Renyi-powerseries}
\end{align}
In Eq.~\eqref{eq:Renyi-powerseries}, the disorder average of the wavefunctions and the eigenvalues can be taken independently. To calculate Eq.~\eqref{eq:Renyi-powerseries} more systematically, one can use a diagrammatic approach introduced in~\cite{Shapourian:2020mkc} and reviewed in appendix~\ref{app:Diag}. After taking the disorder average of the wavefunction, its time dependent part are the following three terms
	\begin{align}
	\tikz[scale=0.7,baseline=0.5ex]{
		\draw[dashed] (0,0) -- (1,0);
		\draw (-0.2,0.)-- (-0.2,-0.25);
		\draw (1.2,-0.1)-- (1.2,0.);
\draw (0.5,0.5) circle (0.2);
\draw (2.5,0.5) circle (0.2);
		\draw[dashed] (2,0) -- (3,0);
		\draw (1.8,0.)-- (1.8,-0.1);
		\draw (3.2,-0.25)-- (3.2,0.);
		\draw (1.2,-0.1)-- (1.8,-0.1);
		\draw (-0.2,-0.25) -- (3.2,-0.25);
		\draw[dashed] (2.0,0) arc (0:180:0.5);
		\draw[dashed] (3.0,0) arc (0:180:1.5);
		\draw (1.8,0) arc (0:180:0.3);
		\draw (3.2,0) arc (0:180:1.7);
	}\quad
	\tikz[scale=0.7,baseline=0.5ex]{
		\draw[dashed] (0,0) -- (1,0);
		\draw (-0.2,0.)-- (-0.2,-0.25);
		\draw (1.2,-0.1)-- (1.2,0.);
\draw (0.5,0.5) circle (0.2);
\draw (2.5,0.5) circle (0.2);
		\draw[dashed] (2,0) -- (3,0);
		\draw (1.8,0.)-- (1.8,-0.1);
		\draw (3.2,-0.25)-- (3.2,0.);
		\draw (1.2,-0.1)-- (1.8,-0.1);
		\draw (-0.2,-0.25) -- (3.2,-0.25);
		\draw[dashed] (2.0,0) arc (0:180:0.5);
		\draw[dashed] (3.0,0) arc (0:180:1.5);
		\draw (1.8,0) arc (0:180:0.3);
		\draw (3.2,0) arc (0:180:1.7);
	}
	\quad&=\frac{d_N^4d_M^2}{d^4}|Z(2it)|^2\,,\\
	\tikz[scale=0.7,baseline=0.5ex]{
		\draw[dashed] (0,0) -- (1,0);
		\draw (-0.2,0.)-- (-0.2,-0.25);
		\draw (1.2,-0.1)-- (1.2,0.);
		\draw[dashed] (2,0) -- (3,0);
		\draw (1.8,0.)-- (1.8,-0.1);
		\draw (3.2,-0.25)-- (3.2,0.);
\draw (0.5,0.25) circle (0.2);
\draw (2.5,0.25) circle (0.2);
		\draw (1.2,-0.1)-- (1.8,-0.1);
		\draw (-0.2,-0.25) -- (3.2,-0.25);
		\draw[dashed] (1.0,0) arc (0:180:0.5);
		\draw[dashed] (3.0,0) arc (0:180:0.5);
		\draw (1.2,0) arc (0:180:0.7);
		\draw (3.2,0) arc (0:180:0.7);
	}\quad
	\tikz[scale=0.7,baseline=0.5ex]{
		\draw[dashed] (0,0) -- (1,0);
		\draw (-0.2,0.)-- (-0.2,-0.25);
		\draw (1.2,-0.1)-- (1.2,0.);
\draw (0.5,0.25) circle (0.2);
\draw (2.5,0.25) circle (0.2);
		\draw[dashed] (2,0) -- (3,0);
		\draw (1.8,0.)-- (1.8,-0.1);
		\draw (3.2,-0.25)-- (3.2,0.);
		\draw (1.2,-0.1)-- (1.8,-0.1);
		\draw (-0.2,-0.25) -- (3.2,-0.25);
		\draw[dashed] (2.0,0) arc (0:180:0.5);
		\draw[dashed] (3.0,0) arc (0:180:1.5);
		\draw (1.8,0) arc (0:180:0.3);
		\draw (3.2,0) arc (0:180:1.7);
	}
	\quad&=\frac{d_N^3d_M^3}{d^4}Z(it)^2Z(-2it)\,,\\
	\tikz[scale=0.7,baseline=0.5ex]{
		\draw[dashed] (0,0) -- (1,0);
		\draw (-0.2,0.)-- (-0.2,-0.25);
		\draw (1.2,-0.1)-- (1.2,0.);
\draw (0.5,0.25) circle (0.2);
\draw (2.5,0.25) circle (0.2);
		\draw[dashed] (2,0) -- (3,0);
		\draw (1.8,0.)-- (1.8,-0.1);
		\draw (3.2,-0.25)-- (3.2,0.);
		\draw (1.2,-0.1)-- (1.8,-0.1);
		\draw (-0.2,-0.25) -- (3.2,-0.25);
		\draw[dashed] (1.0,0) arc (0:180:0.5);
		\draw[dashed] (3.0,0) arc (0:180:0.5);
		\draw (1.2,0) arc (0:180:0.7);
		\draw (3.2,0) arc (0:180:0.7);
	}\quad
	\tikz[scale=0.7,baseline=0.5ex]{
		\draw[dashed] (0,0) -- (1,0);
		\draw (-0.2,0.)-- (-0.2,-0.25);
		\draw (1.2,-0.1)-- (1.2,0.);
\draw (0.5,0.25) circle (0.2);
\draw (2.5,0.25) circle (0.2);
		\draw[dashed] (2,0) -- (3,0);
		\draw (1.8,0.)-- (1.8,-0.1);
		\draw (3.2,-0.25)-- (3.2,0.);
		\draw (1.2,-0.1)-- (1.8,-0.1);
		\draw (-0.2,-0.25) -- (3.2,-0.25);
		\draw[dashed] (1.0,0) arc (0:180:0.5);
		\draw[dashed] (3.0,0) arc (0:180:0.5);
		\draw (1.2,0) arc (0:180:0.7);
		\draw (3.2,0) arc (0:180:0.7);
	}
	\quad&=\frac{d_N^2d_M^4}{d^4}|Z(it)|^4 \,,
	\end{align}
where $d=d_Nd_M$ is the dimension of the Hilbert space of the total system, $d_N$ is the dimension of the Hilbert space of subsystem $A$, and $d_M$ is the dimension of the Hilbert space of subsystem $\bar{A}$. 
In the above diagrams, solid lines and dashed lines represent subsystem $A$ and its complement $\bar{A}$, respectively. In the left three diagrams, each circle represents a factor of $e^{iE_nt}$, while in the right three diagrams, since the matrix elements need to be complex conjugated, each circle represents a factor of $e^{-iE_nt}$. Adding these three terms together, the final result of Eq.~\eqref{eq:Renyi-powerseries} is 
\begin{align}
    {\rm Eq}.~\eqref{eq:Renyi-powerseries}
    =\frac{1}{d^4}
\left(d_N^2d_M^4|Z(it)|^4+d_N^4d_M^2|Z(2it)|^2+2d_N^3d_M^3\Re\left[Z(2it)Z(-it)^2\right]\right) \,.
    \label{eq:Renyi-largeN}
\end{align}
Notice that $|Z(it)|^4$, $|Z(2it)|^2$ and $Z(2it)Z(-it)^2$ are unnormalized, so their magnitudes are of order $d^4=d_N^4d_M^4$, $d^2=d_N^2d_M^2$ and $d^3=d_N^3d_M^3$ respectively. Taking into account the factors in front of each term, the magnitude are of order $d_N^6d_M^8$, $d_N^6d_M^4$, and $d_N^6d_M^6$.
When the dimension of $M$ is sufficiently large, the first term in the parentheses dominates. In this case, the time-dependent part of Eq.~\eqref{eq:Renyi-powerseries} and the second term of Eq.~\eqref{eq:2nd-renyi} are approximately equal. Therefore, the 2nd pseudo-R\'enyi entropy approximates to a constant. When $d_M$ is relatively small, the magnitudes of the last two terms become comparable to that of the first term. In this case, the 2nd pseudo-R\'enyi entropy exhibits the characteristics of the SFF.

\begin{figure}[htbp]
	\centering
	\captionsetup[subfloat]{farskip=10pt,captionskip=1pt}
	\subfloat[t][\centering{N=10}]{\label{SFFSYK4lambda}
		\includegraphics[height =0.265\linewidth]
		{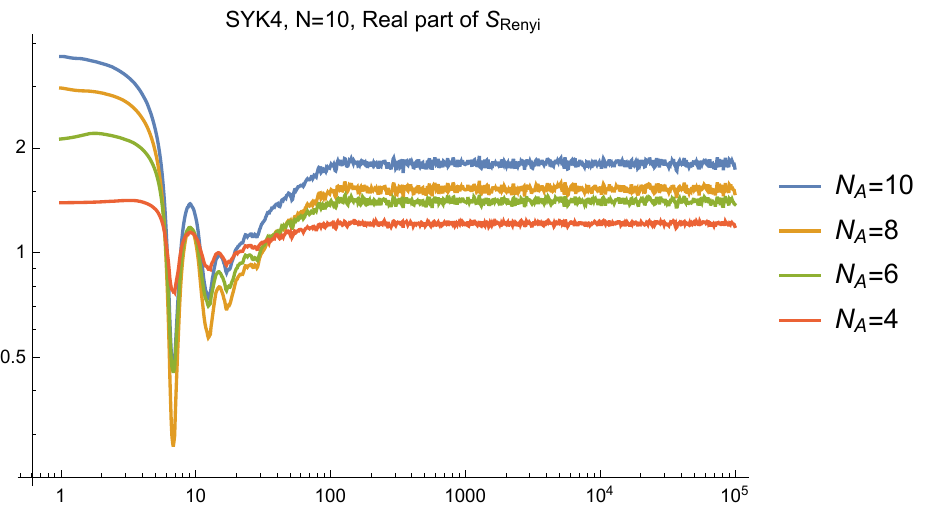}}
	\subfloat[t][\centering{N=12}]{\label{fig:ssyk4_SFF_Lscan_N22}
		\includegraphics[height =0.265\linewidth]
		{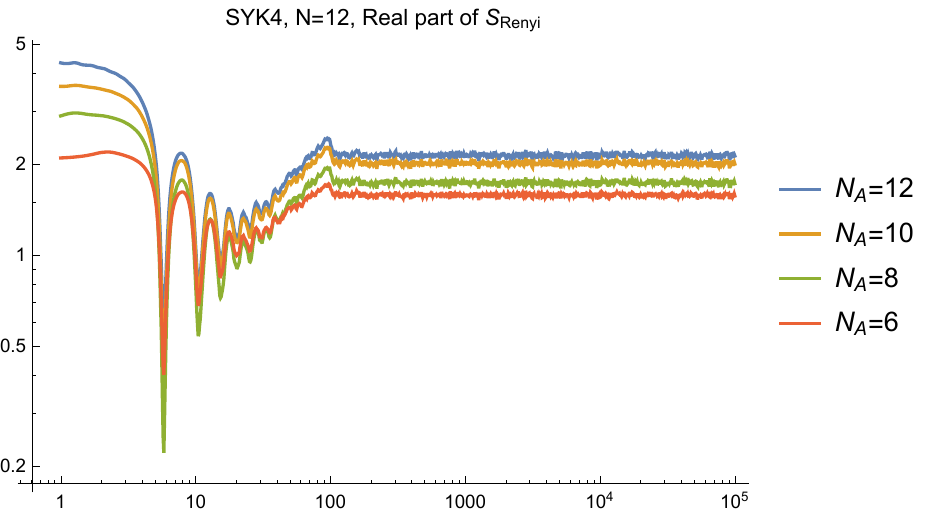}}\\
	\subfloat[t][\centering{N=14}]{\label{SFFSYK4JL}
		\includegraphics[height =0.265\linewidth]
        {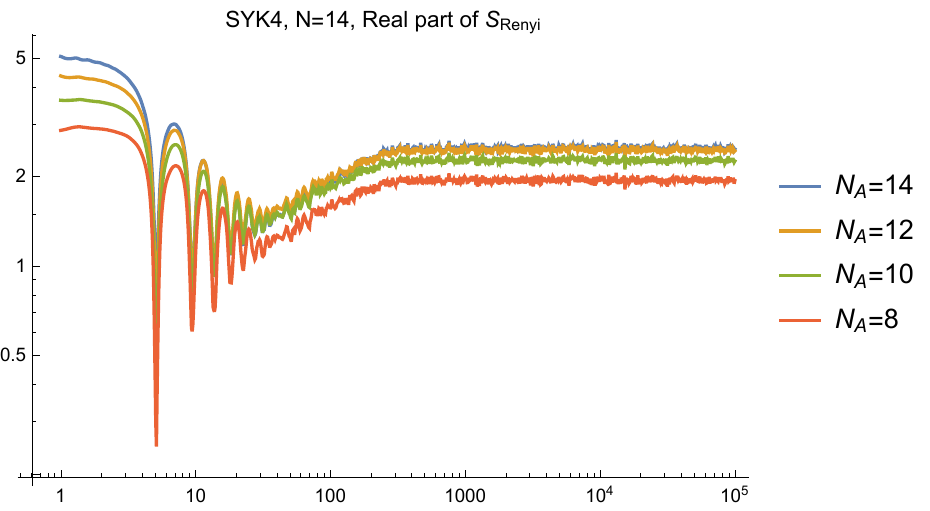}}
	\subfloat[t][\centering{N=16}]{\label{PlotSSYK4SFFcJ0L}
		\includegraphics[height =0.265\linewidth]
        {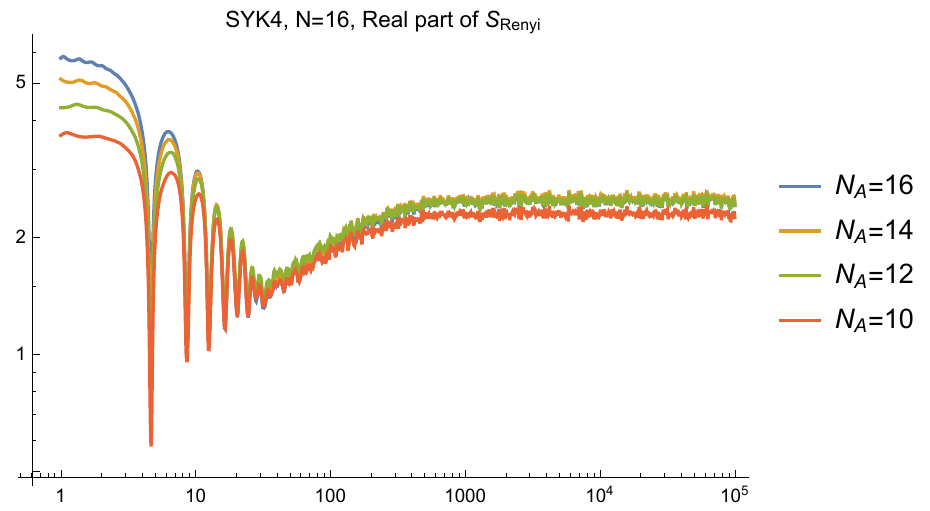}}
	\caption{Plot of the real part of the pseudo-R\'enyi entropy of the SYK$_4$ model with various values of $N$ and $\beta=0$.}
 \label{fig:PEE-SYK4-subregion}
	\vspace{-0.5em}
\end{figure}

For the finite $N$ system, we present the numerical results of the real part of the pseudo-R\'enyi entropy of the subsystem $A$ in Figure~\ref{fig:PEE-SYK4-subregion} with various $N$ and sizes of the subsystem $A$. In Figure~\ref{SFFSYK4lambda}, we observe that when $N_A$ is small, the 2nd pseudo-R\'enyi entropy changes over time small relatively. However, the similar property is not very evident in other figures. For the larger value of $N_A$, the results exhibit the slope-ramp-plateau characteristic observed in the SFF. This property is consistent with the result in Eq.~\eqref{eq:Renyi-largeN}. 

\begin{figure}[htbp]
	\centering
	\captionsetup[subfloat]{farskip=10pt,captionskip=1pt}
	\subfloat[t][\centering{N=16}]{\label{fig:bsSYK4_PRE_kcpscan_N16}
		\includegraphics[height =0.26\linewidth]
		{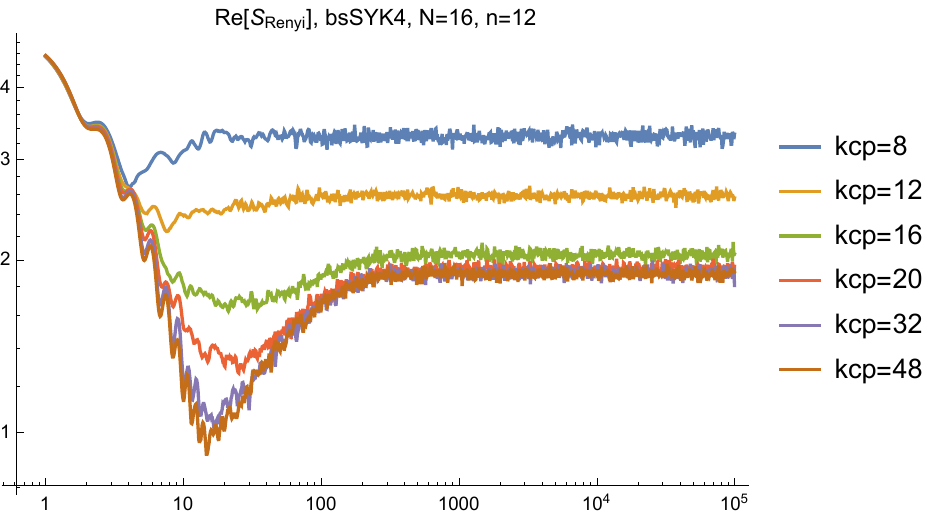}}
	\subfloat[t][\centering{N=18}]{\label{fig:bsSYK4_PRE_kcpscan_N18}
		\includegraphics[height =0.26\linewidth]
		{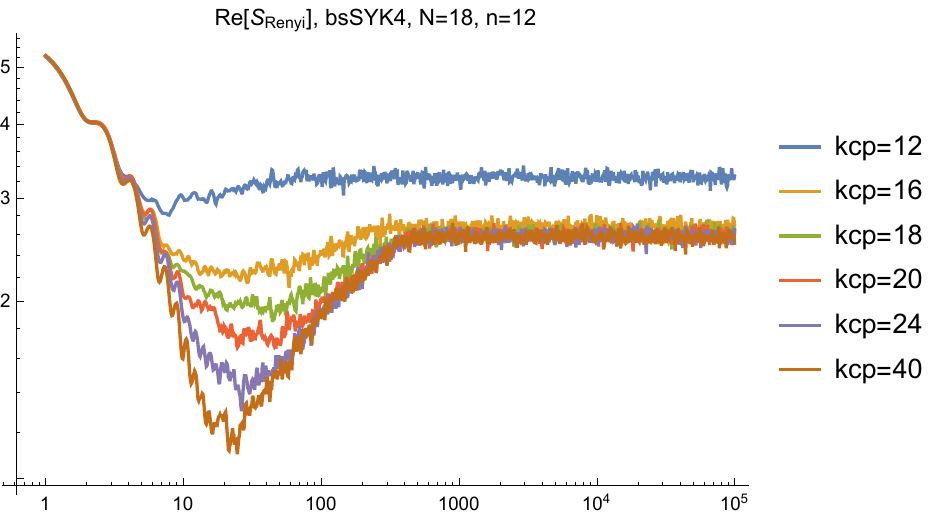}}\\
	\subfloat[t][\centering{N=20}]{\label{fig:bsSYK4_PRE_kcpscan_N20}
		\includegraphics[height =0.26\linewidth]
        {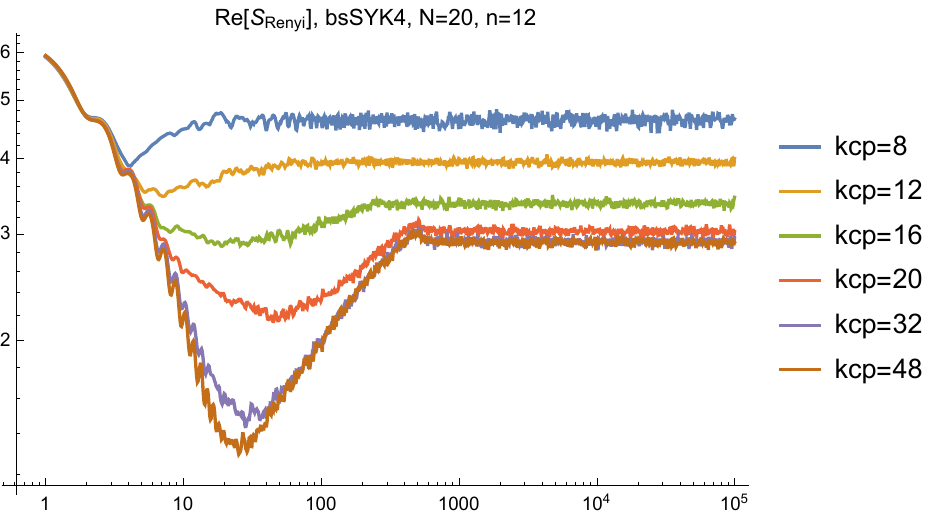}}
	\subfloat[t][\centering{N=22}]{\label{fig:bsSYK4_PRE_kcpscan_N22}
		\includegraphics[height =0.26\linewidth]
        {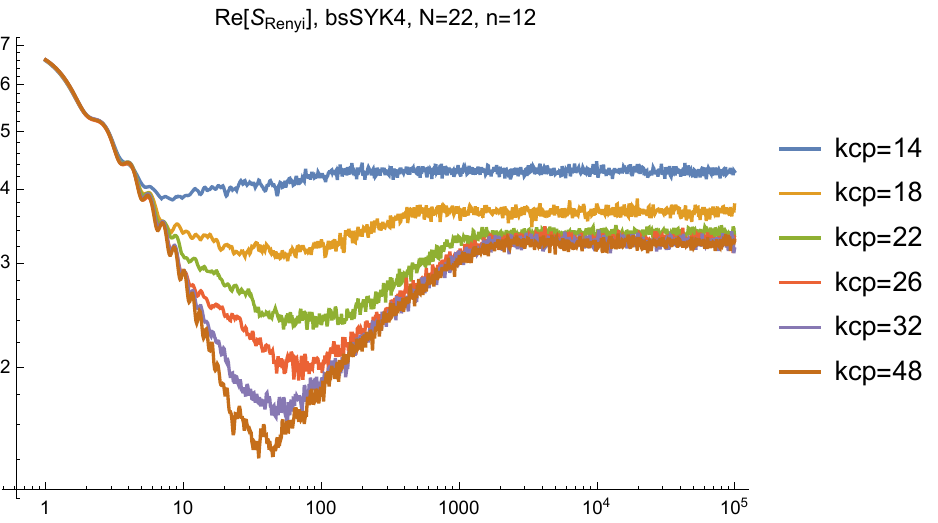}}
	\caption{Real part of the pseudo-R\'enyi entropy of the binary sparse SYK model with different $N$ and $k_{cpl}$.}
 \label{fig:PEE-sparseSYK-Kcpls}
	\vspace{-0.5em}
\end{figure}
\begin{figure}[htbp]
	\centering
	\captionsetup[subfloat]{farskip=10pt,captionskip=1pt}
	\subfloat[t][\centering{N=16}]{\label{fig:XY4_PRE_kcpscan_N16}
		\includegraphics[height =0.26\linewidth]
		{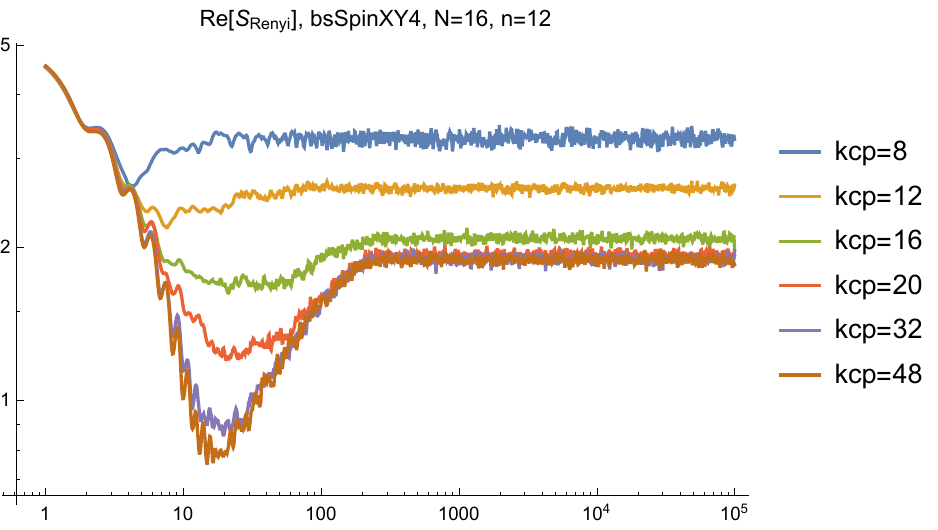}}
	\subfloat[t][\centering{N=18}]{\label{fig:XY4_PRE_kcpscan_N18}
		\includegraphics[height =0.26\linewidth]
		{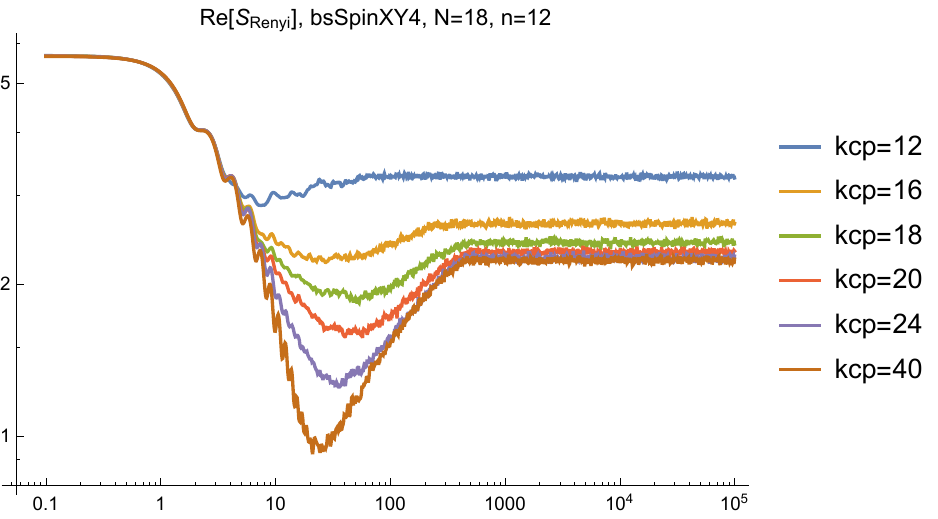}}\\
	\subfloat[t][\centering{N=20}]{\label{fig:XY4_PRE_kcpscan_N20}
		\includegraphics[height =0.26\linewidth]
        {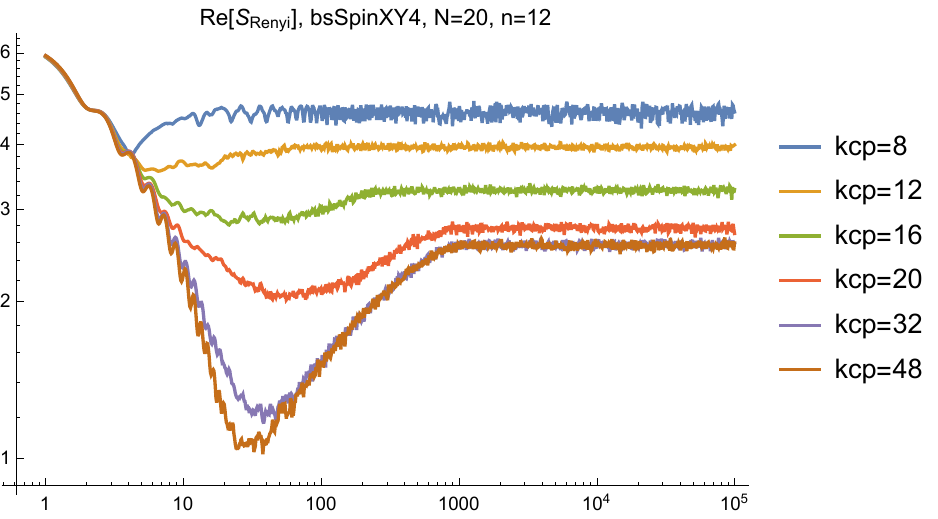}}
	\subfloat[t][\centering{N=22}]{\label{fig:XY4_PRE_kcpscan_N22}
		\includegraphics[height =0.26\linewidth]
        {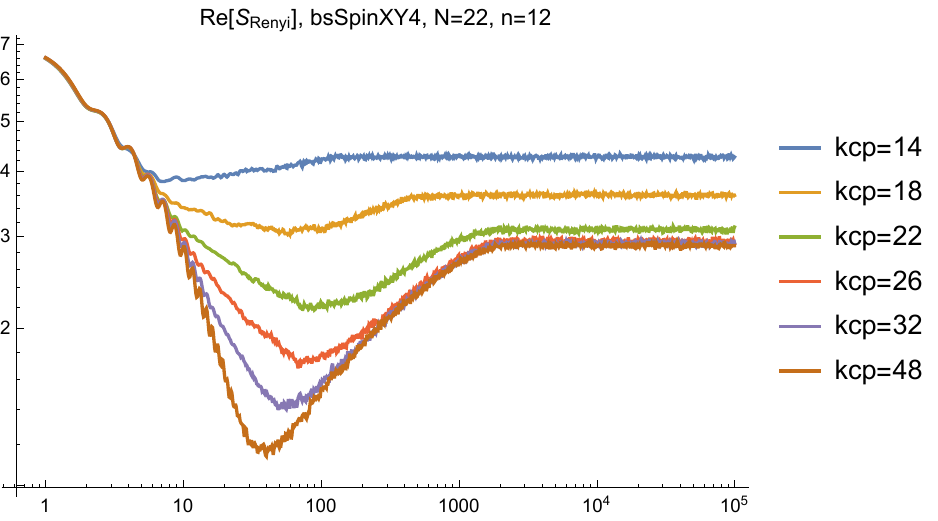}}
	\caption{Real part of the pseudo-R\'enyi entropy of the binary sparse spinXY$_4$ model with different $k_{cpl}$.}
 \label{fig:PEE-sparseSpinXY4-Kcpls}
	\vspace{-0.5em}
\end{figure}

After demonstrating the similarity between the pseudo-R\'enyi entropy of the reduced transition matrix and the SFF, we aim to conduct a more thorough examination of the pseudo-entropy's properties. As discussed in Section~\ref{sec:SYK}, the energy gap ratio of the sparse SYK model undergoes an integrable-chaotic transition as the parameter $k_{cpl}$ increases~\cite{Garcia-Garcia:2020cdo,Caceres:2022kyr,Tezuka:2022mrr}. The critical point of this transition is about $k_{cpl}\sim N$. In the ensuing paragraph, we delve into a numerical exploration of the pseudo-entropy of the sparse SYK model. The outcomes of our investigation are presented in Figure~\ref{fig:PEE-sparseSYK-Kcpls}. Each figure showcases results for a fixed $N$ and varying values of $k_{cpl}$. Across all figures, we discern a consistent pattern: a slope region emerges in the early times, followed by a ramp region during intermediate times, and culminating in a plateau region at later times. Moreover, the time scales $t_d$ and $t_p$ of the sparse SYK model with sufficiently large $k_{cpl}$ closely approximate the values predicted by RMT. Conversely, for the sparse SYK model with lower $k_{cpl}$ values, the time scales $t_d$ and $t_p$ are notably shorter than those of the dense SYK model. From Figure~\ref{fig:PEE-sparseSYK-Kcpls}, one can find this transition occurs around $k_{cpl}\sim N$. Thus, we can confidently conclude that the real part of the pseudo-entropy of the sparse SYK model exhibits an integrable-chaotic transition as the parameter $k_{cpl}$ increases, aligning harmoniously with the findings in~\cite{Tezuka:2022mrr}. In Figure~\ref{fig:PEE-sparseSpinXY4-Kcpls}, we also exhibit the same quantities calculated by the sparse spinXY$_4$ model. We find that the results are consistent with those of the sparse SYK$_4$ model.

\section{The local operator entanglement}
\label{sec:LOE}

In this section, we endeavor to extend the discussions put forth in~\cite{Hosur:2015ylk, Fan:2016ean, Mezei:2016wfz}, to establish a connection between the evolution of pseudo-entropy and the concept of scrambling. It is believed that the scrambling behavior of the evolution operator $U(t)$ can serve as a diagnostic tool for identifying the quantum chaotic properties of certain systems. Moreover, recent work by Dowling et al.~\cite{Dowling:2023hqc} has proven that the linear growth behavior of the R\'enyi entropy is a necessary condition for quantum chaos.

\begin{figure}[htbp]
	\centering
	\captionsetup[subfloat]{farskip=10pt,captionskip=1pt}
	\subfloat[t][\centering{bsSYK$_4$, $\beta=0$}]{\label{fig:PRE-Localquench-N16}
		\includegraphics[height =0.25\linewidth]
		{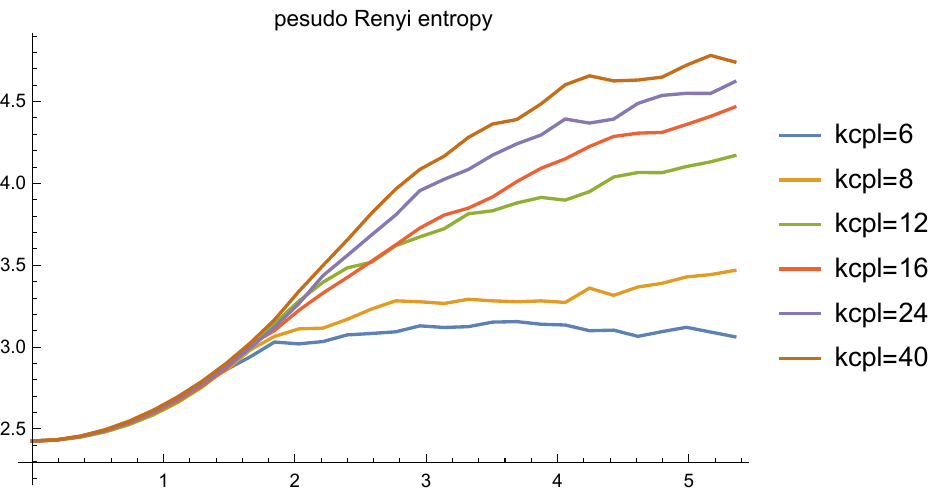}}
	\subfloat[t][\centering{bsSYK$_4$, $\beta=0$}]{\label{fig:OTOC-Localquench-N16}
		\includegraphics[height =0.25\linewidth]
		{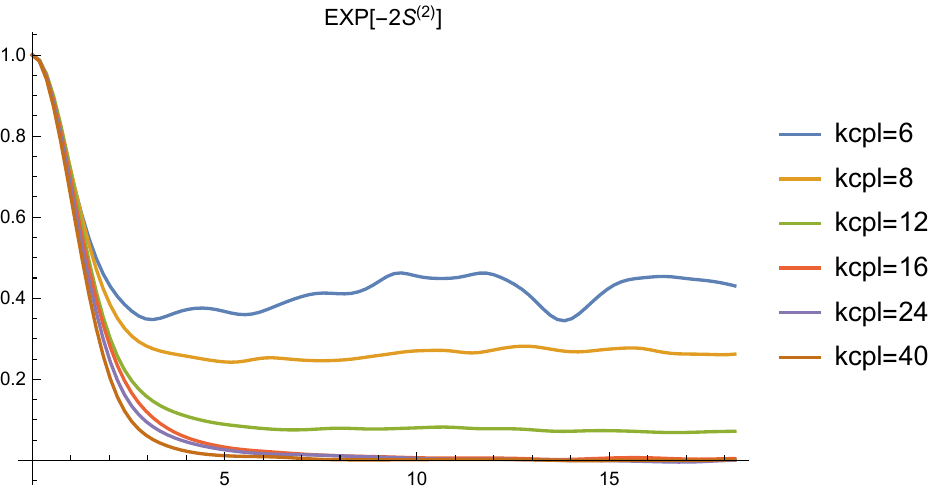}}\\
    \subfloat[t][\centering{bsSpinXY$_4$, $\beta=0$}]{\label{fig:PRE-Localquench-N14}
		\includegraphics[height =0.25\linewidth]
		{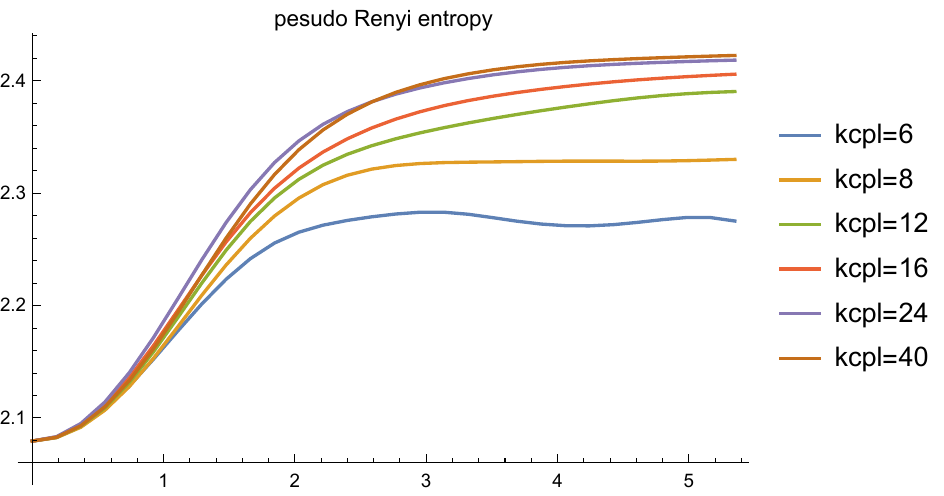}}
	\subfloat[t][\centering{bsSpinXY$_4$, $\beta=0$}]{\label{fig:OTOC-Localquench-N14}
		\includegraphics[height =0.25\linewidth]
		{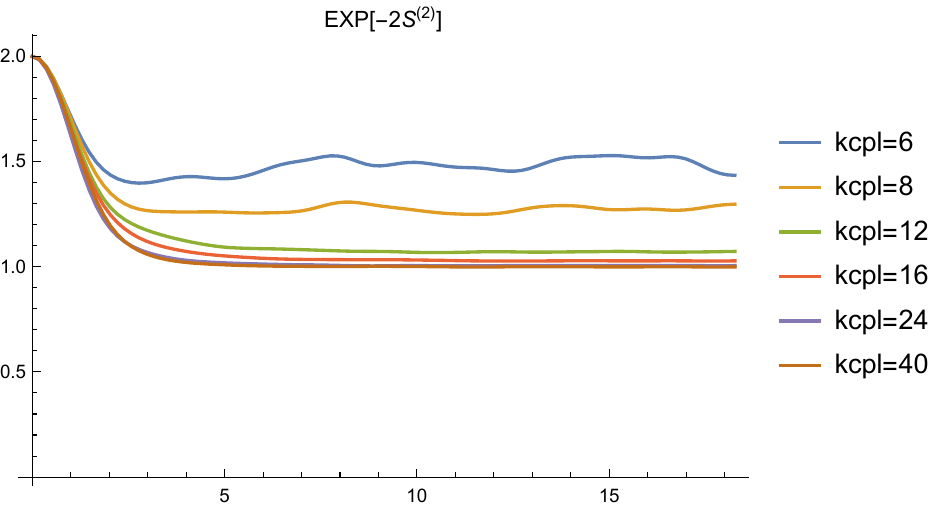}}
	\caption{Plot of the 2nd pseudo-R\'enyi entropy of the binary sparse SYK$_4$ and SpinXY$_4$ model with varous values of $k_{cpl}$. We choose $O_1=\psi_1$ and $O_2=\psi_4$.}
 \label{fig:OTOC-Localquench}
	\vspace{-0.5em}
\end{figure}

The relationship between entanglement entropy and scrambling has been discussed in Ref.~\cite{Hosur:2015ylk,Fan:2016ean}, where they demonstrated that the OTOC is equal to the negative exponential of the second R\'enyi entropy. In order to generalize the R\'enyi entropy to pseudo-R\'enyi entropy, we briefly review the derivation in~\cite{Fan:2016ean}. Consider the 4-point correlation function $\sum_{M\in B}\Tr\left[M(t)O e^{-\beta H}O^\dagger M(t)O e^{-\beta H}O^\dagger\right]$, where the operator $M$ belongs to the subsystem $B$ and satisfies $\sum_{M\in B}MOM=\Tr_B O\otimes I$. Then, the above OTOC can be rewritten as $\Tr_A[\Tr_B[O e^{-\beta H}O^\dagger]\Tr_B[O e^{-\beta H}O^\dagger]]$. In this case, we can define $\rho_A=\Tr_B[O e^{-\beta H}O^\dagger]$ and obtain the following relation 
\begin{align}
    e^{-S_A^{(2)}}=\sum_{M\in B}\Tr\left[M(t)O e^{-\beta H}O^\dagger M(t)O e^{-\beta H}O^\dagger\right] \,.
\end{align}
To generalize the R\'enyi entropy in the above formula to pseudo-R\'enyi entropy, we can naively replace the operator $O e^{-\beta H} O^\dagger$ in the density matrix by $O_1(t_1) e^{-\beta H} O_2(t_2)$. In this case, we can construct the OTOC only by using $O_1$ and $O_2$ without $M$. Considering the normalization factor and purifying the density matrix $e^{-\beta H}$ by a auxiliary system $L$, the transition matrix in this case is
\begin{align}
\mathcal{T}
=\frac{O_1(t_1)|TFD\rangle\langle TFD|O_2^\dagger(t_2)}
{\langle TFD|O_2^\dagger(t_2)O_1(t_1)|TFD\rangle}\,,
\end{align}
which is the transition matrix of the local quenched system~\cite{Guo:2022sfl, He:2023eap}. The reduced transition matrix of the system R is 
\begin{align}
\mathcal{T}_{\rm R}=\Tr_{\rm L}\mathcal{T}
=\frac{O_1(t_1)\rho O_2^\dagger(t_2)}{\Tr\left[O_1(t_1)\rho O_2^\dagger(t_2)\right]}\,,
\end{align}
where $\rho=e^{-\beta H}/\Tr[e^{-\beta H}]$. The exponent of the second pseudo-R\'enyi entropy is
\begin{align}
\Tr\mathcal{T}_R^2
=\frac{\Tr\left[O_1(t_1)\rho O_2^\dagger(t_2)O_1(t_1)\rho O_2^\dagger(t_2)\right]}
{\Tr\left[O_1(t_1)\rho O_2^\dagger(t_2)\right]^2} \,.
\label{eq:OTOC-PEE-localquench}
\end{align}
In contrast to the traditional OTOC, the normalization factor in Eq.~\eqref{eq:OTOC-PEE-localquench} is $\langle O_2^\dagger(t_2)O_1(t_1)\rangle_\beta^2$, which differs from $\langle O_1(\beta)O_1\rangle_\beta\langle O_2(\beta)O_2\rangle_\beta$. It is time-dependent and vanishes at leading order in the large $N$ expansion. In the SYK model, to obtain a standard OTOC, we choose the operators $O_1$ and $O_2$ as follows
\begin{align}
    &O_1=\psi_i(t)+\psi_j(t) \,,\nn\\
    &O_2=\psi_i(t)+\psi_j(0) \,.
\end{align}
At the leading order of $N$, the trace of $\mathcal{T}_R^2$ can be approximated by
\begin{align}
\Tr\mathcal{T}_R^2
&=\frac{\Tr\left[O_1\rho O_2^\dagger O_1\rho O_2^\dagger\right]}
{\Tr\left[O_1\rho O_2^\dagger\right]^2}\nn\\
&=\frac{\Tr\left[\rho^2+\psi_i(t)\rho\psi_j(t)\psi_i(t)\rho\psi_j(t)\right]
+2\Tr\left[\psi_i(t)\rho\psi_j(0)\psi_i(t)\rho\psi_j(0)\right]}
{\Tr\left[\psi_i(t)\rho\psi_i(t)\right]^2}+O(N^{1-q}) \,,
\label{eq:OTOC-bsSYK4}
\end{align}
where the first term in the numerator is time-independent and cancels out precisely when $\beta=0$, while the second term corresponds to a standard OTOC. The infinite temperature OTOC of the SYK model has been obtained in~\cite{Roberts:2018mnp}. Its expression is 
\begin{align}
    F_{\rm OTOC}=1-\frac{1}{2N}\cosh2\mathcal{J}t\,.
\end{align}
The period of the exponential decay occurs between the scales $1/2\mathcal{J}$ and $1/2\mathcal{J}\log 2N$ approximately. At the leading order of the large $N$ expansion, $e^{-2\langle S^{(2)}\rangle_J}$ can be approximated by $\langle e^{-2S^{(2)}}\rangle_J$ where $\langle A\rangle_J$ means the disorder average of $A$. In Figure \ref{fig:OTOC-Localquench}, we provide numerical results for the second pseudo-R\'enyi entropy of the binary sparse SYK model and binary sparse SpinXY$_4$ model. In this figure, we choose $N=16$ and $\mathcal{J}=1/\sqrt{2}$, hence the period of exponential decay is approximately 0.7 to 2.5. As shown in Figure \ref{fig:PRE-Localquench-N16} and \ref{fig:PRE-Localquench-N14}, for larger values of $k_{\text{cpl}}$, there exists a region of linear growth of the second pseudo-R\'enyi entropy around the scrambling time, while for smaller $k_{\text{cpl}}$, this linear growth region is not as prominent. As a comparison, we provide the negative exponent of the second pseudo-R\'enyi entropy, i.e., OTOC in Eq.~\eqref{eq:OTOC-bsSYK4}, in Figure \ref{fig:OTOC-Localquench-N16} and \ref{fig:OTOC-Localquench-N14}. For the SpinXY$_4$ model, the spin operators satisfy commutation relations rather than anti-commutation relations. Therefore, the first term in the numerator of Eq.~\eqref{eq:OTOC-bsSYK4} is not equal to 0, leading to a constant shift of 1 in Figure~\ref{fig:OTOC-Localquench-N14}. After ignoring this shift, the late-time limits of SYK$_4$ and SpinXY$_4$ both tend to be nonzero constants for the smaller values of $k_{\text{cpl}}$, which agrees with the results of Ising CFT obtained in Ref.~\cite{Roberts:2014ifa}.

\section{Summary and prospect}
\label{Sec:Summary}

In this paper, we investigate the possibility of probing the quantum chaotic behavior of a system through pseudo-entropy, an extension of entanglement entropy calculated using the so-called transition matrix $\mathcal{T}$. Unlike the density matrix, the transition matrix can be defined using two non-orthogonal states $|\psi\rangle$ and $|\phi\rangle$, namely $\mathcal{T}=\frac{|\psi\rangle\langle\phi|}{\langle\phi|\psi\rangle}$. By choosing different states $|\psi\rangle$ and $|\phi\rangle$, one can establish connections between pseudo-entropy and various quantum mechanical quantities. This work relates pseudo-entropy and pseudo-R\'enyi entropy to the SFF, OTOC. 

We numerically compute the pseudo-entropy and pseudo-R\'enyi entropy of the SYK$_4$ model using exact diagonalization methods. We present the pseudo-entropy results in Figure \ref{fig:PEE-SYK4-Ns} for different numbers of Majorana fermions, $N$. We have carefully analyzed the reasons for the discrepancy between the pseudo-entropy and the SFF. We found that the reason is from the multi-valued nature (shown in appendix~\ref{sec:log}) of the definition of the pseudo-entropy. Therefore, we instead employ the $n$-th pseudo-R\'enyi entropy to characterize the SFF. In Appendix~\ref{sec:details}, we numerically verify that as $N$ and $n$ increase, the dominant term in the pseudo-R\'enyi entropy approximates the SFF. We display the final results in Figure~\ref{fig:PRE-SYK4}. We further study the subregion pseudo-R\'enyi entropy, showing that it reproduces the SFF shape and remains relatively insensitive to subregion size in Figure \ref{fig:PEE-SYK4-subregion}. Additionally, we investigate the application of pseudo-R\'enyi entropy to distinguish chaotic and integrable systems. By computing the pseudo-R\'enyi entropy in the binary sparse SYK model and the binary sparse SpinXY$_4$ model as a function of $k_{cpl}$ (the number of active Hamiltonian terms), we identify a transition between integrable and chaotic regimes when $k_{cpl}$ exceeds a critical value of ${\cal O}(N)$, as shown in Figures \ref{fig:PEE-sparseSYK-Kcpls} and \ref{fig:PEE-sparseSpinXY4-Kcpls}. This results are consistent with existing research findings~\cite{Roberts:2014ifa,Tezuka:2022mrr,Caceres:2023yoj}. We also investigate the relationship between pseudo-R\'enyi entropy and the out-of-time ordered correlator (OTOC). We find consistent results for two single fermion operators, as demonstrated in Figure \ref{fig:OTOC-Localquench}.

One advantage of the transition matrix is that it can be defined through the weak value of an operator, making it correspond to an observable~\cite{Mollabashi:2021xsd}. Therefore, a natural choice for the states $|\psi\rangle$ and $|\phi\rangle$ is the initial state and final state of the system. 

In Ref.~\cite{Shi:2023czc}, the authors examined how wavefunction $c_{ij}^n$ influences the return probability
\begin{align}
P_{ij,xy}(t) = |\langle xy| e^{-iHt} |ij \rangle|^2 = \sum_{mn} e^{-i(E_m - E_n)t} c_{ij}^m c_{xy}^{m*} c_{ij}^{n*} c_{xy}^n ,,
\end{align}
where $|ij\rangle=|i\rangle\otimes|j\rangle$ and $|xy\rangle=|x\rangle\otimes|y\rangle$ are product states of the subsystems. They demonstrated that for systems with local interactions, the asymptotic behavior of $c_{ij}^n$ follows a Lorentzian distribution, which explains the exponential temporal decay of the return probability. This structure closely resembles the quantities analyzed in Sec.~\ref{sec:PRE-subsystem}, suggesting that pseudo-R\'enyi entropy could similarly serve as a diagnostic tool. However, the models employed in this work—the SYK model and the Pauli SpinXY$_4$ model—lack local interactions, meaning $c_{ij}^n$ does not exhibit this specific behavior. In future studies, we plan to investigate more general systems, including those with local interactions, to further explore the role of pseudo-R\'enyi entropy.

In constructing the connection between pseudo-R\'enyi entropy and OTOCs, we chose a somewhat peculiar state that may not be readily realized by the physical evolution of another state. Therefore, we aim to find a more natural transition matrix to construct OTOCs. 

Additionally, we also aim to establish relationships between pseudo-entropy and other measures of quantum chaos, such as circuit complexity~\cite{Brown:2015bva, Brown:2015lvg, Belin:2021bga} or K-complexity~\cite{Parker:2018yvk, Jian:2020qpp, Rabinovici:2022beu}, among others. Moreover, defining pseudo-negativity based on the mixed-state transition matrix proposed in Eq.~\eqref{eq:TM-mixed} presents a compelling direction for future research.

Finally, we hope to utilize the holographic principle~\cite{Nakata:2020luh} to find the bulk gravitational dual of the transition matrix constructed in the previous sections and establish connections between the properties of pseudo-entropy and the chaotic behavior of black holes.

\section*{Acknowledgments}
We would like to thank Zhuo-Yu Xian, Yu-Xuan Zhang, and Zi-Xuan Zhao for valuable discussions related to this work. S.H. also would like to appreciate the financial support from Jilin University, the Max Planck Partner group, and the Natural Science Foundation of China Grants No.~12475053, No.~12075101, No.~12235016, No.~12347209. P.H.C.L acknowledges the support from JSPS KAKENHI (Grant No.~20H01902 and JP23H01174), and MEXT KAKENHI (Grant No. 21H05462).

\appendix

\section{SFF from pseudo-relative entropy \label{sec:pre-sff}}

Here, we explore the relationship between pseudo-relative entropy and the Spectral Form Factor (SFF). The SFF is a commonly used diagnostic tool for quantum chaos, capable of capturing a system's spectral characteristics through its time evolution. We derive in detail how pseudo-relative entropy reflects a system's quantum chaotic properties via SFF, demonstrating the potential of pseudo-entropy as an effective probe for chaos.

Since entanglement entropy is diverging in QFT, it might be more useful to consider relative entropy, which is a finite quantity even in the continuum limit. The relative entropy $D(\rho|\sigma)$ is defined as
\begin{align}
    D(\rho|\sigma) &\equiv \tr(\rho\log\rho) -\tr(\rho \log \sigma)  \,,
\end{align}
where $\rho$ and $\sigma$ are density matrices with $\rho$ acts as the reference. To connect this quantity to the spectral form factor, we consider the thermofield double state(TFD) of a two-sided model and also its time-evolved state. 
\begin{align}
    &|\psi\rangle=\frac{1}{Z(\beta)^{1/2}}\sum_n e^{-\frac{\beta}{4}(H_L+H_R)}|n_L\rangle\otimes|n_R\rangle \,,\nn \\
    &|\varphi\rangle=e^{\frac{it}{2}(H_L+H_R)}|\psi\rangle\,.
\end{align}
We can then construct the corresponding density matrix and transition matrix using these two states in the form of
\begin{align}
    \rho &= \left|\psi\right\rangle \left\langle \psi \right| \,, \\
    {\cal T}^{\psi|\varphi}&= \frac{\left|\psi\right\rangle \left\langle \varphi \right|}{\left\langle \varphi | \psi \right\rangle} \,,
\end{align}
and then the reduced density matrix and reduced transition matrix of the $R$ system are obtained by partial tracing over the $L$ system
\begin{align}
    \rho_R &= \tr_L \, \rho = \frac{e^{-\beta H_R}}{Z(\beta)} \,, \\
    \mathcal{T}_R^{\psi|\varphi}
    &=\tr_L\mathcal{T}^{\psi|\varphi}
    =\frac{e^{-(\beta+it)H_R}}{Z(\beta+it)} \,.
\end{align}
Using the reduced density matrix and reduced transition matrix, we then compute the generalized relative entropy for these two quantities
\begin{align}
    D\left(\rho_R|\mathcal{T}_R^{\psi|\varphi}\right) &=\tr(\rho_R\log\rho_R) -\tr\left(\rho_R \log \mathcal{T}_R^{\psi|\varphi}\right) \,,\\
    &= it \partial_\beta \log Z(\beta) + \log \frac{Z(t+i\beta)}{Z(\beta)} \,.
\end{align}
With this expression, we can see that the real part of the generalized relative entropy encodes the log of the normalized spectral form factor. We can write down the following expression
\begin{align}
e^{2\Re \left(D\left(\rho_R|\mathcal{T}_R^{\psi|\varphi}\right)\right)} = \left|\frac{Z(t+i\beta)}{Z(\beta)}\right|^2 \,.
\end{align}


\section{Multivaluedness in pseudo-entanglement entropy}
\label{sec:log}

To briefly analyze the impact of the first term of Eq.~\eqref{eq:PE-systemR} on the PEE, we consider the case of the SYK$_q$ model in the large $N$ limit. When $q$ takes different values, the spectral density of the SYK$_q$ model exhibits the following behaviors~\cite{Garcia-Garcia:2016mno,Cotler:2016fpe,Feng:2018zsx}:
\begin{itemize}
\item[1.] For the limit $q^2/N\rightarrow 0$, the spectral density is described by a Gaussian distribution in the interior and a $\sinh$ function near the edge
\begin{align}
\rho(x)=\begin{cases}
\sqrt{\frac{2\pi}{cJ_0^3}}e^{S_0}\sinh(\sqrt{2c(|E_0|-|x|)})\,,\quad &|x|\rightarrow |E_0|\\
2^{N/2-1}\frac{1}{\sqrt{2\pi\sigma}}\exp\left(-\frac{x^2}{2\sigma}\right)\,,&|x|\ll|E_0|
\end{cases}
\end{align}
\item[2.] For the double scaling limit $\mathfrak{q}\equiv e^{-\lambda}\equiv e^{-\frac{q^2}{N}}=\,$const, the spectral density is~\cite{Berkooz:2018jqr}
\begin{align}
\rho(E(\theta))=\frac{\sqrt{1-\mathfrak{q}}}{4\pi\sin\theta}(\mathfrak{q};\mathfrak{q})_\infty
\left|(e^{2i\theta};\mathfrak{q})_\infty\right|^2\,.
\end{align}
where $E(\theta)=\mathcal{J}\frac{2\cos\theta}{\sqrt{\lambda(1-\mathfrak{q})}}$ and $(a;q)_n \equiv \prod_{k=1}^n (1-aq^{k-1})$ is the $q$-Pochhammer symbol.
\item[3.] For the limit $q^2/N\rightarrow\infty$, the spectral density satifies the Wigner semi-circle law
\begin{align}
\rho(x)=\frac{2^{N/2}}{2\pi}\sqrt{4-x^2} \,.
\end{align}
\end{itemize}
For the above three different limits, the results of $\Re\overline{\langle(\beta+it)H\rangle_{\beta+it}}$ are different. For the Schwarzian limit, the result is
\begin{align}
\Re\overline{\langle(\beta+it)H\rangle_{\beta+it}}
\approx\frac{3}{2}-\frac{N\beta\mathcal{J}}{q^2}+\frac{4\pi^2N\alpha_S}{\mathcal{J}}\frac{\beta}{\beta^2+t^2} \,,
\end{align}
where ``$\approx$'' means that we replace $\Re\frac{d}{dt}\langle\ln Z\rangle_J$ by $\Re\frac{d}{dt}\ln\langle Z\rangle_J$. This result is similar to the result for CFT$_2$ in~\cite{Goto:2021kln}. For the semi-circle law limit, the result is
\begin{align}
\Re\overline{\langle(\beta+it)H\rangle_{\beta+it}}\approx
-\Re\frac{2I_2[2(\beta+it)]}{\,_0\tilde{F}_1\left[2,(\beta+it)^2\right]} \,,
\label{eq:H-semicircle}
\end{align}
where $I_n(z)$ is the modified Bessel function of the first kind and ${}_0\tilde{F}_1[a;z]\equiv {}_0F_1[a;z]/\Gamma(a)$ is the regularized hypergeometric function. This result is a rapidly oscillating function with an envelope proportional to $t$.

\begin{figure}[htbp]
	\centering
	\captionsetup[subfloat]{farskip=10pt,captionskip=1pt}
	\subfloat[t][]{\label{fig:H-num-vs-Circle}
		\includegraphics[height =0.25\linewidth]
        {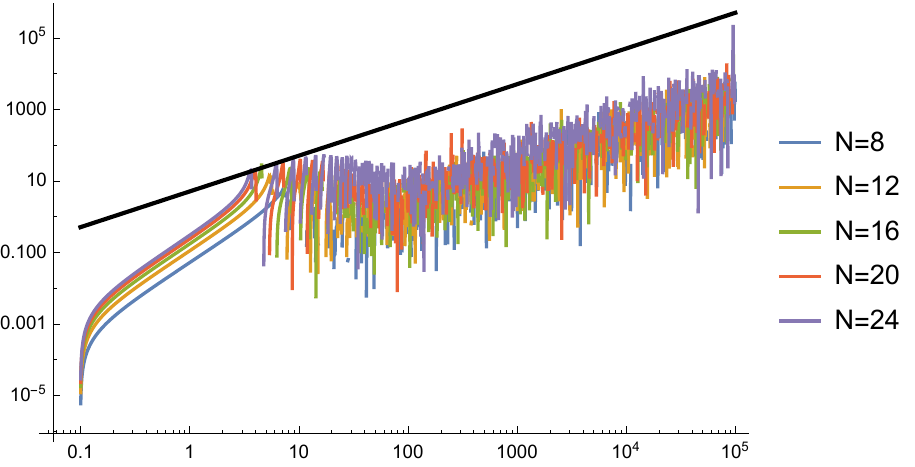}}
	\subfloat[t][]{\label{fig:PEEvsH-SYK4}
		\includegraphics[height =0.25\linewidth]
		{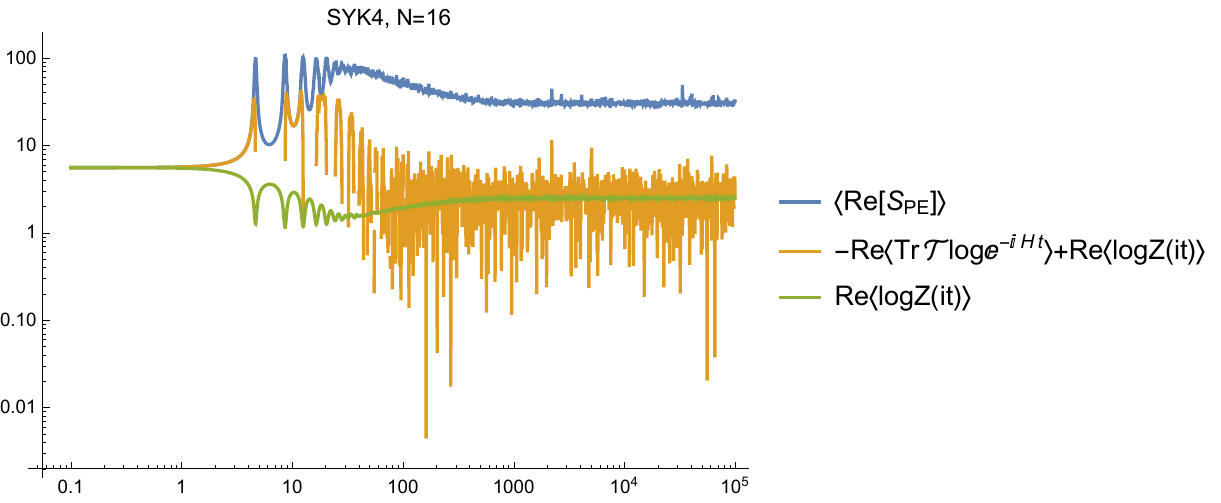}}
	\caption{(a). Plot of $\Re\overline{\langle(\beta+it)H\rangle_{\beta+it}}$ of SYK$_4$ with various $N$ and $\beta=0.1$. The black line is approximate to the envelope of Eq.~\eqref{eq:H-semicircle}, while the colored curves represent the numerical result. The envelope of the colored curves are proportional to $t$ at large $t$. (b). The blue curve and yellow curve correspond to Eq.~\eqref{eq:PEE} and \eqref{eq:lnU-lnZ} respectively. The green curve is $\langle\Re\ln Z(it)\rangle_J$.}
 \label{fig:PEE_multi}
	\vspace{-0.5em}
\end{figure}

For the numerical calculations presented in Figure~\ref{fig:PEE-SYK4-Ns}, all parameters are finite, and the spectral density can be approximated by a Q-Gaussian distribution~\cite{Garcia-Garcia:2017pzl}. However, without taking any further limits, obtaining an explicit expression for the partition function is challenging. Therefore, we numerically computed $\Re\overline{\langle(\beta+it)H\rangle_{\beta+it}}$ and present the results in Figure~\ref{fig:H-num-vs-Circle}. As shown in the figure, the Wigner semicircle distribution provides a good approximation in this scenario.

However, it is clear that this is not the case for Figure~\ref{fig:PEE-SYK4-Ns}, where the late-time behavior is time-independent. We consider that this is due to the ambiguity in the definition of the pseudo-entropy of the transfer matrix $\mathcal{T}_R^{\psi|\varphi}$ in Eq.~\eqref{eq:TM-TFD}. In energy representation, the matrix elements of $\mathcal{T}_R^{\psi|\varphi}$ are complex number $e^{-iE_n t}/Z(it)$. In the calculation of the pseudo-entropy, we need to calculate $\ln[e^{-iE_n t}/Z(it)]$, a logarithm function of a complex number. The logarithm function of a complex number is defined as
\begin{align}
\ln z&=\ln|z|+i\Arg \,z \,, \nn\\
\Arg\left(\frac{z_1}{z_2}\right)&=\Arg \, z_1-\Arg \,z_2\pm 2\pi \,,
\label{eq:arg-sum}
\end{align}
if $\Arg \,z_1-\Arg \,z_2$ is less than $-\pi$ or greater than $\pi$. So we have $\ln e^{i\theta}=i(\theta \text{ mod } 2\pi)\neq i\theta$ and ${\rm Im}\ln e^{i\theta}/z\neq(\theta \text{ mod } 2\pi)+\Arg \,z$. For the pseudo-entropy, the imaginary part of $\ln\mathcal{T}$ will contribute to $\Re \,\tr \,\mathcal{T}\ln\mathcal{T}$ from $\tr\,{\rm Im}\,\mathcal{T}\,{\rm Im}\ln\mathcal{T}$, so we have
\begin{align}
-\Re\,\tr\,\mathcal{T}\ln\mathcal{T}
&=-\Re\,\tr\,\mathcal{T}\ln\left(\frac{e^{-iHt}}{Z(it)}\right)\label{eq:PEE}\\
&\neq -\Re\,\tr\,\mathcal{T}\left(\ln e^{-iHt}-\ln Z(it)\right)\label{eq:lnU-lnZ}\\
&\neq \Re\,\tr\,\mathcal{T}\left(iHt+\ln Z(it)\right)\nn\\
&=\Re\left(it\langle H\rangle_{it}+\ln Z(it)\right)\,.
\label{eq:H-lnZ}
\end{align}
Therefore, the decomposition of pseudo-entropy in Eq.~\eqref{eq:PE-systemR} is not valid\footnote{Physically, considering only the principal value is reasonable because it is evident that $|\psi_1\rangle=\sum_ie^{-(\beta+it)E_i+2i\pi n_i}|i_L,i_R\rangle$ and $|\psi_2\rangle=\sum_ie^{-(\beta+it)E_i}|i_L,i_R\rangle$ describe the same state. Therefore, the pseudo-entropy of the transition matrices $\mathcal{T}_1=|\psi_1\rangle\langle\phi|/\langle\psi_1|\phi\rangle$ and $\mathcal{T}_2=|\psi_2\rangle\langle\phi|/\langle\psi_2|\phi\rangle$ for arbitrary $|\phi\rangle$ should be equal.}. The result in Figure~\ref{fig:H-num-vs-Circle} will contribute to the Eq.~\eqref{eq:H-lnZ} but not to Eq.~\eqref{eq:lnU-lnZ} and \eqref{eq:PEE}. For the pseudo-entropy in the form of Eq.~\eqref{eq:PEE}, if we want to investigate the impact of $e^{-iHt}$ in the log function on the results, we need to calculate the remainder of the difference between each $E_n t$ and $\Arg \,Z(it)$ after dividing by $2\pi$. Providing an analytical computation is nearly impossible; 
therefore, we can only present a comparsion of $\Re[S(\mathcal{T}_R^{\psi|\varphi})]$ and $\Re[S(\mathcal{T}_R^{\psi|\varphi})]-\Re\ln Z(it)$ qualitatively. In the energy representation, the  matrix element of $\mathcal{T}_R^{\psi|\varphi}$ is $e^{-i E_n t}/Z(it)$. For simplicity, we can denote it is as $\rho e^{i\theta}=\frac{e^{i\theta_1}}{\rho^{-1}e^{i\theta_2}}$ where $\theta=\theta_1-\theta_2$ and $Z=\rho^{-1}e^{i\theta_2}$. Using these notations, the real part of the matrix element of $-\Re\,\mathcal{T}_R^{\psi|\varphi}\ln\mathcal{T}_R^{\psi|\varphi}$ and $-\Re\,\mathcal{T}_R^{\psi|\varphi}(\ln e^{-iHt}-\ln Z)$ can be expressed as
\begin{align}
-\Re\,\mathcal{T}_R^{\psi|\varphi}\ln\mathcal{T}_R^{\psi|\varphi}
:&-\rho\cos\theta\ln\rho+\rho\sin\theta \,\Arg \, e^{i(\theta_1-\theta_2)}\nn\\
-\Re \, \mathcal{T}_R^{\psi|\varphi}(\ln e^{-iHt}-\ln Z)
:&-\rho\cos\theta\ln\rho+\rho\sin\theta(\Arg \, e^{i\theta_1}-\Arg \, e^{i\theta_2})
\end{align}
According to Eq.~\eqref{eq:arg-sum}, the matrix element of $(-\Re \,\mathcal{T}_R^{\psi|\varphi}\ln\mathcal{T}_R^{\psi|\varphi})-(-\Re\,\mathcal{T}_R^{\psi|\varphi}(\ln e^{-iHt}-\ln Z))$ is
\begin{align}
\begin{cases}
-2\pi\rho\sin\theta\,,\quad &\pi<\Arg \, e^{i\theta_1}-\Arg \, e^{i\theta_2}<2\pi  \,, \\
2\pi\rho\sin\theta\,,\quad &-2\pi<\Arg \,e^{i\theta_1}-\Arg \, e^{i\theta_2}<-\pi \,.
\end{cases}
\end{align}
For the first case, $\sin\theta<0$, while for the second case, $\sin\theta>0$. Therefore, for any $\theta$, $\theta_1$ and $\theta_2$, this difference is always positive. This implies that for each matrix element in the trace and for any $t$ this difference is positive. In other words, Eq.~\eqref{eq:PEE} is always greater than Eq.~\eqref{eq:lnU-lnZ}, as illustrated in Figure~\ref{fig:PEEvsH-SYK4}. From Figure~\ref{fig:PEEvsH-SYK4}, one can find that in the ramp and plateau regions, the effect of $e^{-iHt}$ in log function contributes to the $\Re \,[S(\mathcal{T}_R^{\psi|\varphi})]$ significantly\footnote{In a recent work~\cite{Caputa:2024gve}, Caputa et al. also investigated the pseudo-entropy of the transfer matrix~\eqref{eq:TM-TFD}. They calculated the pseudo-entropy using Eq.~\eqref{eq:H-lnZ} and take the time-averaged result. In our calculation, since the results of Eqs.~\eqref{eq:lnU-lnZ} and~\eqref{eq:H-lnZ} are rapidly oscillating functions of time at late time, it is reasonable to assume that the time-averaged values of their first term are zero. However, as indicated in Figure~\ref{fig:PEEvsH-SYK4}, taking the time average still does not make Eqs.~\eqref{eq:PEE} (blue line) and \eqref{eq:lnU-lnZ} (yellow line) equal.}.


\section{Commutativity of logarithms and the ensemble averages}
\label{sec:details}

\begin{figure}[!htb]
		\centering
		\includegraphics[width=1\linewidth]{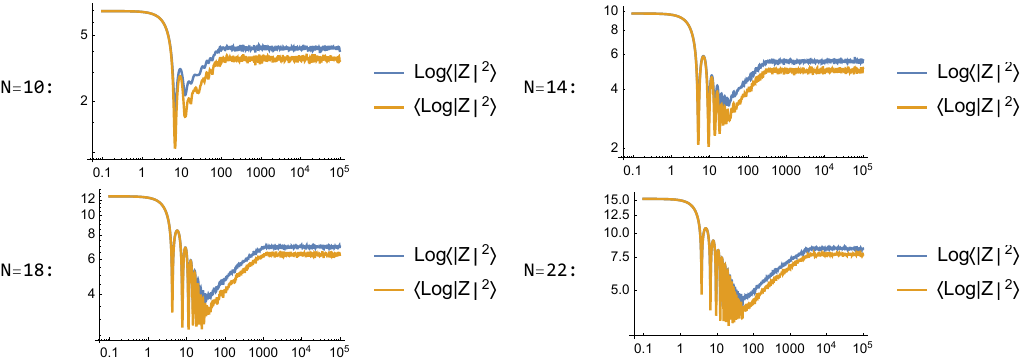}
	\caption{The comparison of ${\rm log}\left\langle |Z|^2\right\rangle_J$, and $\left\langle {\rm log}|Z|^2\right\rangle_J$ with different values of $N$ and $\beta=0$.}
	\label{fig:Z2-SYK4-Ns}
\end{figure}

In Section~\ref{sec:SFF}, we explored the possibility of characterizing the SFF using pseudo-entanglement entropy and pseudo-Rényi entropy. This approach requires interchanging the order of the logarithmic operation and ensemble averaging. In this appendix, we provide numerical verification of this issue. Figure~\ref{fig:Z2-SYK4-Ns} demonstrates the difference between $\log\langle|Z|^2\rangle_J$ and $\left\langle {\rm log}|Z|^2\right\rangle_J$ for the second term in Eq.~\eqref{eq:PE-systemR}.

\begin{figure}[!htb]
		\centering
		\includegraphics[width=1\linewidth]{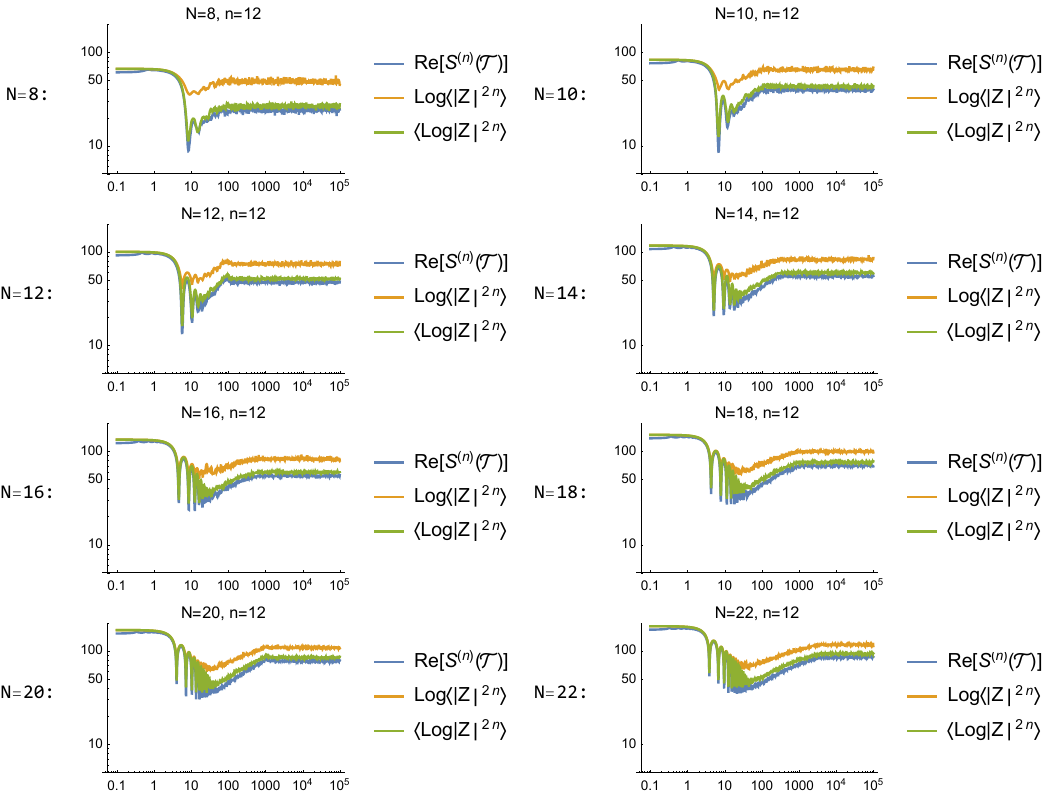}
	\caption{The comparison of ${\rm Re}[S^{(n)}\mathcal{T}]$, ${\rm log}\left\langle |Z|^n\right\rangle_J$, and $\left\langle {\rm log}|Z|^n\right\rangle_J$ with different values of $N$ and $\beta=0$.}
	\label{fig:Renyi-Zn-SYK4-Ns}
\end{figure}
\begin{figure}[!htb]
		\centering
		\includegraphics[width=1\linewidth]{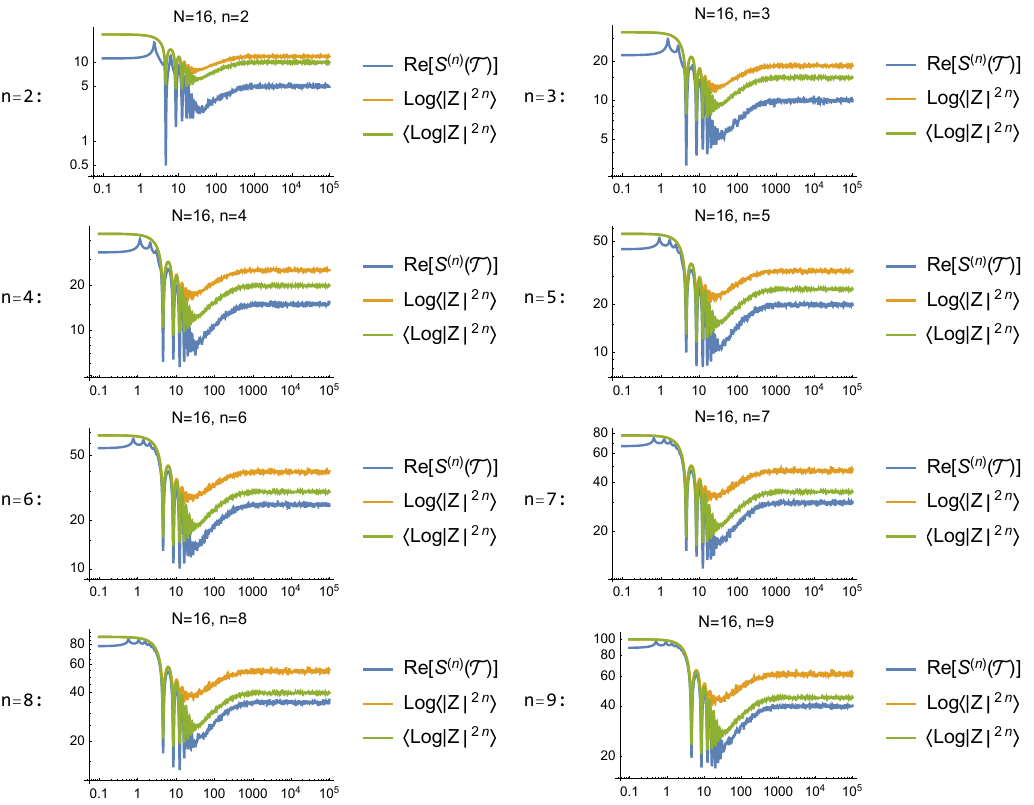}
	\caption{The comparison of $(2n-1){\rm Re}[S^{(n)}\mathcal{T}]$, ${\rm log}\langle |Z|^n\rangle_J$, and $\langle {\rm log}|Z|^n\rangle_J$ of the SYK$_4$ model with different values of $n$ and $\beta=0$.}
	\label{fig:Renyi-Zn-SYK4-ks}
\end{figure}

In Figure~\ref{fig:Renyi-Zn-SYK4-Ns} and Figure~\ref{fig:Renyi-Zn-SYK4-ks}, we provide the numerical comparison of the real part of the $n$-th pseudo-R\'enyi entropy, $\left\langle\log|Z|^{2n}\right\rangle_J$ and $\log\left\langle|Z|^{2n}\right\rangle_J$, appeared in Eq.~\eqref{eq:renyiPEE}, for different $N$ and different $n$ respectively. We observe that the curves of the $n$-th pseudo-R\'enyi entropy approach the curves of $\left\langle\log|Z(\beta+it)|^{2n}\right\rangle_J$ as $n$ increases. 

\begin{figure}[!htb]
		\centering
		\includegraphics[width=0.6\linewidth]{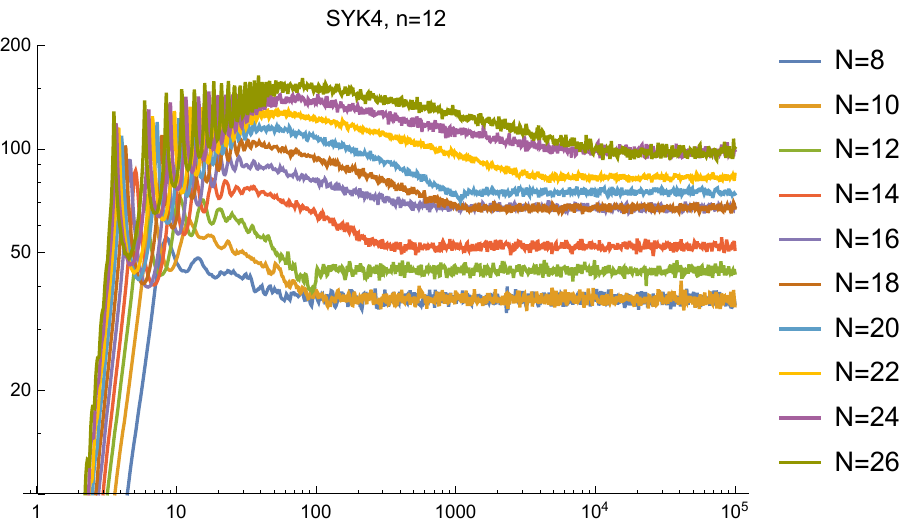}
	\caption{Plot of $2(1-n)\Re\left[S^{(n)}(\mathcal{T}_R^{\psi|\varphi})\right]+(n-1)N\ln 2$ for various values of $N$ and $\beta=0$.}
	\label{fig:normalizedRenyiPE-SYK4-Ns}
\end{figure}

In Figure~\ref{fig:normalizedRenyiPE-SYK4-Ns}, we examine whether pseudo-R\'enyi entropy can reflect the system's symmetry properties.
When using the pseudo-R\'enyi entropy to characterize the SFF, the process involves interchanging the logarithm operation with the averaging operation, leading to a slight discrepancy between the logarithm of SFF and the pseudo-R\'enyi entropy, as shown in Figures~\ref{fig:Renyi-Zn-SYK4-Ns} and \ref{fig:Renyi-Zn-SYK4-ks}. Thus, we aim to investigate whether the pseudo-R\'enyi entropy can also correctly reflect the energy level degeneracy. To characterize the degeneracy of the models, we include the pseudo-R\'enyi entropy of the SYK model with $N$ Majorana fermions multiplied by the normalization constant $(n-1)N\ln2$. The modified expression is given by
 \begin{align}
     2(1-n)\Re\left[S^{(n)}(\mathcal{T}_R^{\psi|\varphi})\right]+(n-1)N\ln2
     &=\left\langle\log\frac{\left|Z(int)\right|^2}{{d_N}^2}\right\rangle_J
     -\left\langle\log\frac{\left|Z(it)\right|^{2n}}{{d_N}^{2n}}\right\rangle_J\,,
\label{eq:renormalizedRenyi-SYK4-Ns}
 \end{align}
where $d_N=2^{N/2}=Z(0)$ represents the dimension of the Hilbert space of the SYK model with $N$ Majorana fermions. The results for various values of $N$ are displayed in Figure~\ref{fig:normalizedRenyiPE-SYK4-Ns}. The negative logarithmic term in Eq.~\eqref{eq:renormalizedRenyi-SYK4-Ns} contributes to the growth behavior in the slope region and the decay behavior in the ramp region. We observe that the values of $N$ mod 8 = 0 and $N$ mod 8 = 2 coincide in the plateau region, reflecting the double degeneracy of the SYK model with $N$ mod 8 = 2, 4, 6, while being nondegenerate for $N$ mod 8 = 0.


\section{Diagrammatic approach \label{app:Diag}}
	In this appendix, we will introduce a diagrammatic approach proposed in~\cite{Shapourian:2020mkc} to calculate the pseudo-R\'enyi entropy in Section~\ref{sec:SFF}. Consider a system in the state $\left|\Psi\right\rangle$. We divide it into two subsystems $A$ and $B$, with the state denoted as
\begin{align}
\left|\Psi\right\rangle=\sum_{i=1}^{d_A}\sum_{a=1}^{d_B}X_{ia}\left|\Psi_A^i\right\rangle\otimes\left|\Psi_B^a\right\rangle\,,
\end{align}
where $d_A$ ($d_B$) is the dimension of the subsystem $A$ $(B)$. The element of the density matrix corresponding to this pure state is $X_{ia}X^*_{bj}$ and can be expressed diagrammatically as 
	\begin{align}
	\rho_{ia,jb}=|\Psi\rangle\langle\Psi|_{ia,jb}=
	\tikz[baseline=-0.5ex]{
		\draw[dashed] (0,0.2) node[align=center, above] {\footnotesize $a$} -- (0,-0.2);
		\draw[dashed] (1,0.2) node[align=center, above] {\footnotesize $b$} -- (1,-0.2);
		\draw (-0.2,0.2) node[align=center, above] {\footnotesize $i$} -- (-0.2,-0.15);
		\draw (1.2,0.2) node[align=center, above] {\footnotesize $j$} -- (1.2,-0.15);
	}\ ,
\end{align}
where solid lines and dashed lines correspond to subsystem $A$ and $B$, respectively. The matrix product is represented by connecting the bottom of the two lines
\begin{align}
\label{eq:tr_r2}
    (\rho^2)_{ia,jb}= 
    \,
    \tikz[baseline=0ex]{
    \draw[dashed] (0,0.2) node[align=center, above] {\footnotesize $a$} -- (0,-0.25);
    \draw[dashed] (1,-0.25) -- (2,-0.25);
    \draw[dashed]  (1,0.2) -- (1,-0.25);
    \draw (-0.2,0.2) node[align=center, above] {\footnotesize $i$} -- (-0.2,-0.25);
    \draw (1.2,-0.1)-- (1.2,0.2);
    \draw[dashed] (2,0.2)-- (2,-0.25);
    \draw[dashed] (3,0.2) node[align=center, above] {\footnotesize $b$} -- (3,-0.25);
    \draw (1.8,0.2)-- (1.8,-0.1);
    \draw (3.2,0.2) node[align=center, above] {\footnotesize $j$} -- (3.2,-0.25);
    \draw (1.2,-0.1)-- (1.8,-0.1);
    }\ ,
\end{align}
The reduced density matrix is represented by connecting the bottom of the two lines in the same matrix 
\begin{align}
    \label{eq:rho_diag}
    [\rho_A]_{i,j}= 
    {\sum_{a=1}^{d_B} X_{ia} X_{aj}^*}
= 
\,
    \tikz[baseline=-0.5ex]{
    \draw[dashed] (0,0.2) node[align=center, above] {\footnotesize $a$} -- (0,0);
    \draw[dashed] (0,0)  -- (1,0);
    \draw[dashed]  (1,0.2) node[align=center, above] {\footnotesize $a$} -- (1,0);
    \draw (-0.2,0.2) node[align=center, above] {\footnotesize $i$} -- (-0.2,-0.15);
    \draw (1.2,0.2) node[align=center, above] {\footnotesize $j$} -- (1.2,-0.15);
    }\ .
\end{align}
The average operation is represented by connecting the two lines overhead
\begin{align}
\tikz[baseline=-0.1ex,scale=0.9]{
    \draw[dashed] (1.0,0) arc (0:180:0.5);
    \draw (1.2,0.) arc (0:180:0.7);
    }
    \ \,
    := \langle X_{ia}X_{ bj}^\ast \rangle=  \frac{1}{d_A d_B}\, \delta_{ij} \delta_{ab} \,.
    \label{eq:doubleline1}
\end{align}
The ensemble average of each pair of random variables contributes a factor of $1/d = 1/(d_Ad_B)$, and each closed loop of solid (dashed) lines contributes a factor of $d_A$ ($d_B$).

In this work, we need to replace the density matrix with the transition matrix $\mathcal{T}=e^{-(\beta+it)H}|\Psi\rangle\langle\Psi|$, so we need to generalize the diagrammatic approach described above. The element of the transition matrix mentioned above is
\begin{align}
\mathcal{T}_{ia,jb}=\sum_{n=1}^d e^{-(\beta+it)E_n}X^n_{ia}{X^n_{bj}}^* \,.
\label{eqn:matrixT}
\end{align}
Therefore, for each matrix element, we add a factor of $e^{-(\beta+it)E_n}$ labeled by a circle as
	\begin{align}
	\mathcal{T}_{ia,jb}=
	\tikz[baseline=-0.5ex]{
		\draw[dashed] (0,0.2) node[align=center, above] {\footnotesize $a$} -- (0,-0.2);
		\draw[dashed] (1,0.2) node[align=center, above] {\footnotesize $b$} -- (1,-0.2);
        \draw (0.5,0) circle  (0.2); 
		\draw (-0.2,0.2) node[align=center, above] {\footnotesize $i$} -- (-0.2,-0.15);
		\draw (1.2,0.2) node[align=center, above] {\footnotesize $j$} -- (1.2,-0.15);
	}\ ,
\end{align}
and the rule of the ensemble average of each pair of random variables becomes
\begin{align}
\langle X^m_{ia}{X^n_{ bj}}^\ast \rangle=  \frac{1}{d_A d_B}\, \delta_{ij} \delta_{ab}\delta^{mn}\,.
\end{align}
The other rules remain unchanged. With this diagrammatic technique, we evaluate Eq. (\ref{eq:Renyi-powerseries}) and obtain the result Eq. (\ref{eq:Renyi-largeN}). 

	\bibliographystyle{JHEP}
	
	\bibliography{ref}
 
\end{document}